\documentclass[fleqn,usenatbib]{mnras}
\usepackage{newtxtext,newtxmath}

\usepackage[T1]{fontenc}

\DeclareRobustCommand{\VAN}[3]{#2}
\let\VANthebibliography\thebibliography
\def\thebibliography{\DeclareRobustCommand{\VAN}[3]{##3}\VANthebibliography}

\usepackage{natbib}
\usepackage{xspace}
\usepackage{amsmath}
\usepackage{lipsum}
\usepackage{appendix}
\usepackage{multirow}
\usepackage{bm}
\usepackage{graphicx}	

\newcommand{\orcid}[1]{}

\newcommand{\msun}{$M_{\odot}$\xspace}
\newcommand{\rsun}{$R_{\odot}$\xspace}
\newcommand{\lsun}{$\mathrm{L_{\odot}}$\xspace}

\newcommand{\mstar}{\ensuremath{M_{\star}}\xspace}
\newcommand{\rstar}{\ensuremath{R_{\star}}\xspace}

\newcommand{\rhostar}{\ensuremath{\rho_{\star}}\xspace}

\newcommand{\teff}{\ensuremath{T_{\mathrm{eff}}}\xspace}

\newcommand{\logg}{\ensuremath{\log g}\xspace}
\newcommand{\feh}{\ensuremath{[\mbox{Fe}/\mbox{H}]}\xspace}

\newcommand{\gcc}{g\,cm$^{-3}$\xspace}

\newcommand{\emcee}{{\texttt{emcee}}\xspace}
\newcommand{\batman}{{\texttt{batman}}\xspace}

\newcommand{\celerite}{{\texttt{celerite}}\xspace}
\newcommand{\ldtk}{{\texttt{PyLDTk}}\xspace}
\newcommand{\platon}{{\texttt{PLATON}}\xspace}
\newcommand{\corner}{{\texttt{corner}}\xspace}

\newcommand{\abc}{Astrobiology Center, 2-21-1 Osawa, Mitaka, Tokyo 181-8588, Japan}
\newcommand{\naoj}{National Astronomical Observatory of Japan, 2-21-1 Osawa, Mitaka, Tokyo 181-8588, Japan}
\newcommand{\hongo}{Department of Astronomy,  Graduate School of Science, The University of Tokyo, 7-3-1 Hongo, Bunkyo, Tokyo 113-0033, Japan}
\newcommand{\sokendai}{Department of Astronomy, School of Science, The Graduate University for Advanced Studies (SOKENDAI), 2-21-1 Osawa, Mitaka, Tokyo, Japan}

\newcommand{\komaba}{Department of Multi-Disciplinary Sciences, Graduate School of Arts and Sciences, The University of Tokyo, 3-8-1 Komaba, Meguro, Tokyo 153-8902, Japan}
\newcommand{\iac}{Instituto de Astrof\'\i sica de Canarias (IAC), 38205 La Laguna, Tenerife, Spain}
\newcommand{\laguna}{Departamento de Astrof\'{i}sica, Universidad de La Laguna (ULL), 38206 La Laguna, Tenerife, Spain}
\newcommand{\komabasc}{Komaba Institute for Science, The University of Tokyo, 3-8-1 Komaba, Meguro, Tokyo 153-8902, Japan}

\newcommand{\steff}{$3207 \pm 58$\xspace}
\newcommand{\slogg}{$4.944 \pm 0.031$\xspace}
\newcommand{\sfeh}{$0.15 \pm 0.03$\xspace}
\newcommand{\smass}{$0.2634 \pm 0.0077$\xspace}
\newcommand{\srad}{$0.2932 \pm 0.0093$\xspace}
\newcommand{\srho}{$14.7 \pm 1.5$\xspace}
\newcommand{\sdist}{$45.01_{-0.17}^{+0.16}$\xspace}
\newcommand{\slumi}{$0.00816 \pm 0.00029$\xspace}
\newcommand{\upperT}{\xspace190}
\newcommand{\upperf}{\xspace61}

\title[Characterization of starspots on K2-25]{Characterization of starspots on a young M-dwarf K2-25: multi-band observations of stellar photometric variability and planetary transits}

\author[M. Mori et al.]{
Mayuko Mori \orcid{0000-0003-1368-6593},$^{1}$\thanks{E-mail: mayukomori.519@gmail.com}
Kai Ikuta \orcid{0000-0002-5978-057X},$^{1}$
Akihiko Fukui \orcid{0000-0002-4909-5763},$^{2,3}$
Norio Narita \orcid{0000-0001-8511-2981},$^{2,3,4}$
Jerome P. de Leon \orcid{0000-0002-6424-3410},$^{1}$
\newauthor{
John H. Livingston \orcid{0000-0002-4881-3620},$^{4,5,6}$
Masahiro Ikoma \orcid{0000-0002-5658-5971},$^{7}$
Yugo Kawai \orcid{0000-0002-0488-6297},$^{1}$
Kiyoe Kawauchi \orcid{0000-0003-1205-5108},$^{8}$ 
}
\newauthor{
Felipe Murgas \orcid{0000-0001-9087-1245},$^{3,9}$ 
Enric Palle \orcid{0000-0003-0987-1593},$^{3,9}$ 
Hannu Parviainen \orcid{0000-0001-5519-1391},$^{3,9}$ 
Gareb Fern\'{a}ndez Rodr\'{i}guez \orcid{0000-0003-0597-7809},$^{3,9}$ 
}
\newauthor{
Yuka Terada \orcid{0000-0003-2887-6381},$^{10,11}$ 
Noriharu Watanabe \orcid{0000-0002-7522-8195}, $^{1}$ 
Motohide Tamura \orcid{0000-0002-6510-0681}$^{4,5,12}$ 
}
\\ \\
$^{1}$\komaba\\
$^{2}$\komabasc\\
$^{3}$\iac\\
$^{4}$\abc\\
$^{5}$\naoj\\
$^{6}$\sokendai\\
$^{7}$Division of Science, National Astronomical Observatory of Japan, 2-21-1 Osawa, Mitaka, Tokyo 181-8588, Japan \\
$^{8}$Department of Physical Sciences, Ritsumeikan University, Kusatsu, Shiga 525-8577, Japan\\
$^{9}$\laguna\\
$^{10}$Institute of Astronomy and Astrophysics, Academia Sinica, P.O. Box 23-141, Taipei 10617, Taiwan, R.O.C.\\
$^{11}$Department of Astrophysics, National Taiwan University, Taipei 10617, Taiwan, R.O.C.\\
$^{12}$\hongo\\
}

\date{Accepted 2024 March 20. Received 2024 March 19; in original form 2024 February 12}

\pubyear{2024}

\begin{document}
\label{firstpage}
\pagerange{\pageref{firstpage}--\pageref{lastpage}}
\maketitle

\begin{abstract}
Detailed atmospheric characterization of exoplanets by transmission spectroscopy requires careful consideration of stellar surface inhomogeneities induced by starspots. This effect is particularly problematic for planetary systems around M-dwarfs, and their spot properties are not fully understood.
We investigated the stellar activity of the young M-dwarf K2-25 and its effect on transit observations of the sub-Neptune K2-25\,b. From multi-band monitoring observations of stellar brightness variability using ground-based telescopes and TESS, we found that the temperature difference between the spots and photosphere is $<$\upperT\,K and the spot covering fraction is $<$\upperf\% ($2\sigma$).
We also investigated the effect of starspot activity using multi-epoch, multi-band transit observations. 
We rule out cases with extremely low spot temperatures and large spot covering fractions. The results suggest that spots could distort the transmission spectrum of K2-25\,b by as much as $\sim100$\,ppm amplitude, corresponding to the precision of JWST/NIRSPEC of the target.
Our study demonstrates that simultaneous multi-band observations with current instruments can constrain the spot properties of M-dwarfs with good enough precision to support atmospheric studies of young M-dwarf planets via transmission spectroscopy.
\end{abstract}


\graphicspath{{./}{figures/}}

\section{Introduction} \label{sec:intro}

The number of discovered exoplanets now exceeds 5000, largely due to space-based transit surveys, such as the Kepler space telescope \citep{Borucki2010} and the Transiting Exoplanets Survey Satellite \citep[TESS;][]{Ricker2015}.
Among the various planetary groups, planets feasible for transmission spectroscopy are one of the most interesting targets to characterize their atmospheres. The James Webb Space Telescope \citep[JWST;][]{JWST2006} and future space telescopes such as Ariel \citep{Tinetti2018, Ariel2021}, with their extremely high precision, enable us to detect atmospheric signals of less than 100\,ppm, which allows more detailed analysis of the planetary atmospheric compositions.

Recently, the transit light source effect (TLSE) has been identified as a significant factor affecting planetary transmission spectra\citep{Rackham2018,Rackham2023}. In brief, unocculted spots or faculae during planetary transits make the observed transit depth shallower or deeper than the expected transit depth without spots or faculae. This effect is particularly complicated for the planets around active M-dwarfs: the wavelength dependence of this effect is confounded by the absorption of molecules present in both stellar photosphere and spots, resulting in false positive signals of planetary atmospheres. Some studies have investigated the possibility that signals in transmission spectra can be interpreted as either due to planetary atmospheres or TLSE \citep[e.g.,][]{Edwards2021, Gressier2022, Barclay2021, Moran2023}.

In atmospheric retrievals of exoplanet spectra, it is common practice to make empirical assumptions about the stellar surface models, which are specified in spot temperatures and covering fractions (the ratio of projected spot area to the stellar disk) \citep[e.g.,][]{Pinhas2018, Cracchiolo2021}. In other cases, it is common to optimize transmission spectra by atmosphere models and stellar surface models simultaneously \citep[e.g.,][]{Lim2023, Fournier-Tondreau2023}. However, many degeneracies between stellar surface and planetary atmospheric models make it challenging to assume/retrieve realistic stellar surface models \citep{Rackham2023b,Berardo2023}. In addition, what makes the assumption of stellar surface models difficult is that the spot characteristics have not been well studied, especially for M-dwarfs. The commonly used empirical relations of spot temperature and stellar surface temperature \citep{Berdyugina2005, Maehara2017, Herbst2021} are derived mainly from observations of solar-type stars. 

A useful method to constrain spot temperature is multi-band photometric observations of rotational modulation of stellar brightness caused by spots appearing and disappearing in the observed hemisphere of the star \citep[e.g.,][]{Henry1995,Morris2018}.
In this method, it is necessary to perform simultaneous multi-band observations because the spot distribution could be different if two observations are widely separated in time. However, there are a few studies with such observational strategies \citep{Rosich2020} because intensive observations are required. 

The purpose of this study is to investigate the spot activity of an M-dwarf and its impact on planetary transits through multi-band observations in the optical. As a target, we selected the active M-dwarf K2-25, which harbours a sub-Neptune (K2-25\,b) that is a candidate for future atmospheric characterization.
To derive its current spot characteristics, we performed simultaneous multi-band monitoring observations of K2-25 with ground-based telescopes concurrently with TESS. In addition, we conducted ground-based multi-band transit observations of K2-25\,b to assess the effects of these spot characteristics on measurements of planetary transits. 

\subsection{K2-25 system} \label{sec:k2_25}

\begin{table}
\scriptsize
\centering
\caption{Fundamental parameters of K2-25. The source ``Gaia DR3" means the data from the Gaia mission Data Release 3 \citep{Gaia2016, GaiaDR3_2023}.}
\label{tab:stellar_K2-25}
\begin{tabular}{lrr}
\hline
Parameter & Value & Source \\
\hline
\multicolumn{3}{l}{\it Equatorial coordinates, parallax, and proper motion}  \\
\noalign{\smallskip}
R.A. (J2000)	& 04$^\mathrm{h}$13$^\mathrm{m}$05.613$^\mathrm{s}$ & Gaia DR3 \\ 
Dec. (J2000)	& $+$15$\degr$14$\arcmin$52.02$\arcsec$	& Gaia DR3 \\
$\pi$ (mas) 	& $22.3572 \pm 0.0308$ & Gaia DR3 \\
$\mu_\alpha$ (mas\,yr$^{-1}$) 	& $122.450 \pm 0.038$		& Gaia DR3 \\
$\mu_\delta$ (mas\,yr$^{-1}$) 	& $-18.603 \pm 0.026$		& Gaia DR3 \\
\hline
\multicolumn{3}{l}{\it Fundamental parameters}   \\
\teff (K) & \steff & \citet{Thao2020} \\
\feh (dex) & \sfeh & \citet{Mann2016}  \\
\mstar (\msun) & \smass & \citet{Thao2020}  \\
\rstar (\rsun) & \srad & \citet{Thao2020}  \\
\rhostar (\gcc) & \srho & \citet{Thao2020}  \\
\logg (cgs) & \slogg & \citet{Thao2020} \\
distance (pc) & \sdist & \citet{Thao2020}  \\
Luminosity (\lsun) & \slumi & \citet{Thao2020}  \\
Age (Gyr) & $0.730_{-0.052}^{+0.050}$ & \citet{Mann2016}\\
\hline
\end{tabular}
\end{table}

K2-25 is a young M4.5 dwarf in the Hyades cluster. Its sub-Neptune-sized planet K2-25\,b was discovered by Kepler space telescope in K2 Campaign 4 \citep{Mann2016}. The fundamental parameters of the K2-25 system are summarized in Table~\ref{tab:stellar_K2-25}.

As K2-25 is a member of the Hyades cluster, the age of the system is well-constrained to be $650\pm70$\,Myr \citep{Martin2018}. This makes possible an in-depth discussion of the formation and evolution process of the planet's atmosphere, once observational constraints on the atmosphere are obtained. With its relatively bright host star and deep transits, the planet is a promising target for atmospheric studies with the JWST, Ariel, and next-generation telescopes. 

However, the star has a brightness variability with a period of $\sim 1.88$ days, suggesting that it is an active star with spots. As they would affect the result of transmission spectroscopy, several previous studies have investigated the atmospheric properties of K2-25\,b with the stellar activity of the host star in mind. 

\citet{Thao2020} derived transit depths from a low-resolution transmission spectrum with multiple transit datasets obtained from several space- and ground-based telescopes. 
They compared the derived spectrum with atmosphere models, combined with stellar surface models assuming a spot temperature of 2800\,K or 3000\,K on a stellar surface of 3200\,K. They found that the transmission spectrum of K2-25\,b prefers a flat model to a clear solar-abundance atmosphere model, as a large spot covering fraction (22\% or 36\%) was needed to fit a solar-abundance atmosphere model, and such a large spot covering fraction contradicts the observed amplitude of brightness variations. Therefore, they concluded that, although TLSE could be at play to some extent, the atmospheric spectrum of K2-25\,b is flat, which indicates a cloudy/hazy atmosphere and/or a high mean molecular weight atmosphere.

\citet{Gaidos2020} and \citet{Stefansson2020} measured the stellar obliquity and planetary mass via spectroscopic observations. Their radial velocity observations yielded a planetary mass higher than that expected from a mass-radius relation, making the amplitude of the atmospheric signal in the transmission spectrum smaller without necessitating the presence of clouds or hazes.

No detailed space-based transmission spectroscopy has ever been performed for K2-25, and such observations by JWST and similar instruments are needed to investigate whether the atmospheric spectrum of K2-25\,b is flat. On the other hand, high-precision spectroscopy will need to pay more attention to TLSE, as it will be sensitive to small differences in spot properties, which can complicate the interpretation of atmospheric models. \citet{Thao2020} made empirical assumptions about spot temperatures, but more realistic stellar surface models can be constrained by observations. Interest in the K2-25 system itself and the need to understand its current spot activity are the reasons we chose this system as our target, in addition to the fact that it was a good test case for our methods.

\vskip\baselineskip
The remainder of the paper is organized as follows: Section~\ref{sec:observations} summarizes the observational data used in our analyses; Section~\ref{sec:analyses} describes our analysis methods for constraining the spot characteristics from multi-band monitoring and transit light curves; Section~\ref{sec:discussion} describes K2-25's spot characteristics and the effect of the spots on the transit observations, and compares our results with previous studies. Section~\ref{sec:coclusion} summarizes the results and discussions, with the future prospects of studying spot properties of M-dwarfs.


\section{Observations \& Data Reductions}\label{sec:observations}
The observations are divided into two categories: monitoring photometry to investigate stellar brightness variations using TESS, Sinistro, and ZTF (Section \ref{sec:tess}, \ref{sec:sinistro}, and \ref{sec:ztf}); transit photometry to evaluate the effect of spots on transit depth measurements by TESS, MuSCAT2, and MuSCAT3 (Section \ref{sec:tess} and \ref{sec:muscats}). These observations were conducted at optical wavelengths, where the spot effect is expected to be prominent. The data are summarized in Table~\ref{tab:observations}.

\subsection{Monitoring and transit photometry - TESS} \label{sec:tess}
K2-25 (TIC 434226736, TOI-5095) was observed by TESS in Sector 44 (2021 October 12 to November 6) during TESS Cycle 4 with a 2-min cadence in the TESS bandpass (600–1000 nm). We used the {\tt PDCSAP} (Pre-search Data Conditioning Simple Aperture Photometery) light curves produced by the Science Processing Operations Center (SPOC) photometry pipeline \citep{Jenkins2002, Jenkins2010, Jenkins2016} for both the monitoring and transit analyses. We also checked the {\tt SAP} (Simple Aperture Photometry-) light curves, as sometimes stellar modulation is removed from the {\tt PDCSAP} light curves. We confirmed that the brightness variation on K2-25 was almost identical in the {\tt SAP} and {\tt PDCSAP} light curves, both in shape and normalized amplitude, with instrumental systematics and a long-term trend corrected in {\tt PDCSAP} light curve. In addition, while nearby stars blending into the TESS aperture may affect photometric accuracy, we confirmed that their effects were negligible (Appendix~\ref{sec:ap-aperture}).

To analyse the brightness variability, we removed outliers with more than $3\sigma$ away from the median of the normalized light curve. In the transit analyses, for the six transits observed by TESS, non-detrended light curves were cut out for a range of 0.1 days before and after the predicted transit-centre time.

\subsection{Monitoring photometry - LCO 1m / Sinistro} \label{sec:sinistro}

Sinistro is a single-band optical camera mounted on five 1m telescopes worldwide, operated by Las Cumbres Observatory \citep[LCO,][]{LCOGT2013}. Our observations were performed specifically with the telescopes at McDonald Observatory in Texas and at Teide Observatory in Tenerife, Canary islands. Sinistro has a $26'.5 \times 26'.5$ field of view with a pixel scale of 0.389\arcsec pix$^{-1}$. 

We observed K2-25 from 2021 October 02 to November 05 with Sinistro in $g$-band (400–550 nm), and from 2021 October 13 to 2022 January 31 in $z$-band (820-920 nm). With the exposure time of 120 seconds in $g$-band and 40 seconds in $z_s$-band (denoted $z$-band hereafter for simplicity), 5-15 frames were taken every 4-5 hours during good observing conditions with airmass $<$ 1.6, moon distance $> 30$ deg, and no weather or instrumental issues. Data reduction was carried out through the LCO \texttt{BANZAI} pipeline \citep{McCully2018}.
The photometry was obtained using the package \texttt{astropy-Photutils} \citep{photutils}. We found that circular apertures with a radius of 8 pixels were optimal to reduce the scatter in the light curves. We also subtracted the sky background noise calculated using an annulus with a radius of 50 to 80 pixels. 

We adopted the differential photometry method to account for the telluric variability. In practice, the telluric effect on stellar brightness variations should depend on the colour of the star, given that atmospheric extinction occurs to different degrees for stars of different colours. To minimise this effect, we only used stars with colours similar to the target as comparison stars. We calculated the $g-z$ colour of each star in the field of view from the catalogue of Pan-STARRS Data Release 2 \citep{PS1, PS2}. As the $g-z$ value of K2-25 is 3.48, we selected stars with $g-z> 2$ for use in the calculation. 

We excluded frames with the following three conditions as outliers: (i) the number of detected stars was smaller than 50; (ii) the median stellar magnitude was $> 0.7$ mag fainter than that of the reference frame; (iii) the obtained target brightness was more than $3\sigma$ different from the mean model obtained by fitting a third-order polynomial.

\subsection{Monitoring photometry - ZTF} \label{sec:ztf}

The Zwicky Transient Facility (ZTF) is a time-domain survey that conducts a full sweep of the northern visible sky every two days \citep{ztf1}. The survey is performed by a camera with a 47 square degree field of view mounted on the Samuel Oschin 1.2m telescope at Palomar Observatory in California \citep{ztf1}. 

We used the archival photometry of K2-25 in $Z_g$ and $Z_r$-bands  (367.6-561.4 and 549.8-739.4 nm), which is similar to SDSS $g$ and $r$-band \citep{ztf2}, from 2018 March 27 until 2022 March 2.
For the ZTF light curves, we removed the data points with the quality warning flags (expressed as `catflags' in the ZTF catalogue), and data points where zero-magnitude is 3$\sigma$ away from the median value of zero-magnitude. In addition, the magnitude was corrected for the second-order extinction effect by a linear function of colour coefficients (`clrcoeff' in the ZTF catalogue). Finally, the magnitude was converted to flux, from which we produced a normalised light curve.

We found that the shape of brightness variations of K2-25 had changed over a timescale of $\sim 200$ days possibly due to the variation of spot distribution (see Appendix~\ref{sec:an-longterm}). Therefore, in order to maintain simultaneous observations with the other dataset, we only used the data after MJD $=$59400 (2021 July 5) in the analyses.

\subsection{Transit photometry - MuSCAT2 and MuSCAT3} \label{sec:muscats}

MuSCAT2 and MuSCAT3 are simultaneous multi-band cameras with four optical channels of $g$, $r$, $i$, and $z$- bands (400–550, 550–700, 700–820, and 820–920 nm) \citep{Narita2019, Narita2020}. MuSCAT2 is installed on the 1.52m Telescopio Carlos S\'{a}nchez in the Teide Observatory, Canaries, and MuSCAT3 is on 2m Faulkes Telescope North at Las Cumbres Observatory (LCO) on Haleakala, Maui. MuSCAT2 and MuSCAT3 share common features, while they are different in pixel scale (0.44\arcsec and 0.27\arcsec for MuSCAT2 and 3) and field of view ($7'.4 \times 7'.4$ and $9'.1 \times 9'.1$). 

We observed four transits of K2-25\,b with MuSCAT2 on 2021 December 8, 22, 29, and 2022 January 5, and two transits with MuSCAT3 on 2022 January 8 and 29. The exposure times and number of data for each observing night are summarised in Table~\ref{tab:observations}. A transit number $N_{\rm T}$ was introduced to indicate how many transits have occurred with the transit at $T_0=2458515.64206$ (BJD) \citep{Stefansson2020} as the starting point. We note that the $i$-band camera was not available on 2021 December 8 and 22, because of technical difficulties with the instrument. 

These observations were not exactly simultaneous with the TESS observations, and there was a gap of 35 days ($\sim 10$ planetary orbital periods) between the last transit with TESS and the first transit with MuSCAT2.

The reduction for MuSCAT2 data and the aperture photometry of both MuSCAT2/3 datasets were conducted by the custom pipeline described in \citet{Fukui2011}, and the reduction for MuSCAT3 data was carried out by the \texttt{BANZAI} pipeline.
For each night and each band, the optimal aperture radius was found by selecting the value between 8 and 14 pixels that minimized the scatter in the light curve. Between one and three stars brighter than the target were selected as comparison stars for each night and each band.

The derived raw light curves are shown in Figure~\ref{fig:outliers}. Apart from the transit signal, some rapid flux increases were detected especially in shorter wavelengths, which are likely to be flares. 
In particular, we visually detected a significant flare-like slope in $g$ and $r$-band light curves for transit $N_{\rm T} = 307$. We also detected a flux bump during the transit for light curve $N_{\rm T} = 314$, which might be a flare during the transit or a spot crossing.

The removal of outliers was conducted as follows. First, we removed a flux jump caused by a flare-like signal in $N_{\rm T} = 307$. As the highest peak of the signal was at BJD$=2459585.36556$, the data points 0.01\,day before and after this timestamp were removed. Although the flare-like signal was not visually obvious in $i$ and $z$-band, we removed the points in all bands to reduce the bias from arbitrary manipulation. Then, we fit each light curve at each band and each date one by one, using mean models as the transit model combined with the baseline model. The transit model was calculated by the package \batman \citep{BATMAN} from the transit parameters described in Section~\ref{sec:an-transit}. The baseline model was defined by a third-order polynomial of time. The models were optimized using MCMC package \emcee \citep{emcee}. We removed outliers with more than $3\sigma$ away from the mean model. We repeated this operation three times (Figure~\ref{fig:outliers}).

\begin{table*}
\centering
\caption{Summary of the observational data used in the analyses, where $N_{\rm T}$ indicates the number of transits counted from the reference transit-centre time $T_0$.}
\begin{tabular}{lccccc}\hline
\multicolumn{4}{l}{\bf{Monitoring photometry}} \\ 
Date        & Instrument & Filters      &Exposure Times (sec) & \\ \hline
2021 Oct 12 - Nov 06 & TESS & TESS & 120 & \\
2021 Oct 02 - Nov 05 & Sinistro & SDSS $g$ & 120 & \\
2021 Oct 13 - 2022 Jan 31 & Sinistro & Pan-STARRS $z_s$ & 40 & \\ 
2021 Jul 05 - 2022 Mar 02 & ZTF & $Z_g, Z_r$ & 30 & \\  \hline
\multicolumn{4}{l}{\bf{Transit photometry}}& \\ 
Date        & Instrument & Filters      &Exposure Times (sec) &  $N_{\rm T}$ \\ \hline
2021 Dec 08 & MuSCAT2    & SDSS $g, r, z_s$    & 60, 45, 15 & 299  \\
2021 Dec 22 & MuSCAT2    & SDSS $g, r, z_s$    & 60, 45, 15 & 303  \\
2021 Dec 29 & MuSCAT2    & SDSS $g, r, i, z_s$ & 60, 45, 30, 15  &305  \\
2022 Jan 05 & MuSCAT2    & SDSS $g, r, i, z_s$ & 60, 45, 30, 15 &307  \\
2022 Jan 08 & MuSCAT3    & SDSS $g, r, i, z_s$ & 120, 50, 23, 15 &308  \\
2022 Jan 29 & MuSCAT3    & SDSS $g, r, i, z_s$ & 150, 45, 23, 15 &314  \\ \hline
\end{tabular}
\label{tab:observations}
\end{table*}

\section{Analyses \& Results}\label{sec:analyses}
The analyses described in this section are divided into two main parts. First is the analysis of stellar brightness variability to constrain the spot characteristics and to calculate the degree of TLSE (Section~\ref{sec:an-rotation} to \ref{sec:dis-tlse-stack}). Second is the multi-band transit depth derivation to evaluate the TLSE (Section~\ref{sec:an-transit}). Additional analyses and results are described in the Appendix.

\subsection{Monitoring light curve analyses - rotation period}\label{sec:an-rotation}
We obtained five light curves: the TESS light curve, Sinistro light curves in $g$- and $z$-band, and ZTF light curves in $Z_g$- and $Z_r$-band. These light curves allow us to constrain the temperature and distribution of spots on K2-25.

\begin{figure}
  \centering
      \includegraphics[width=0.4\textwidth]{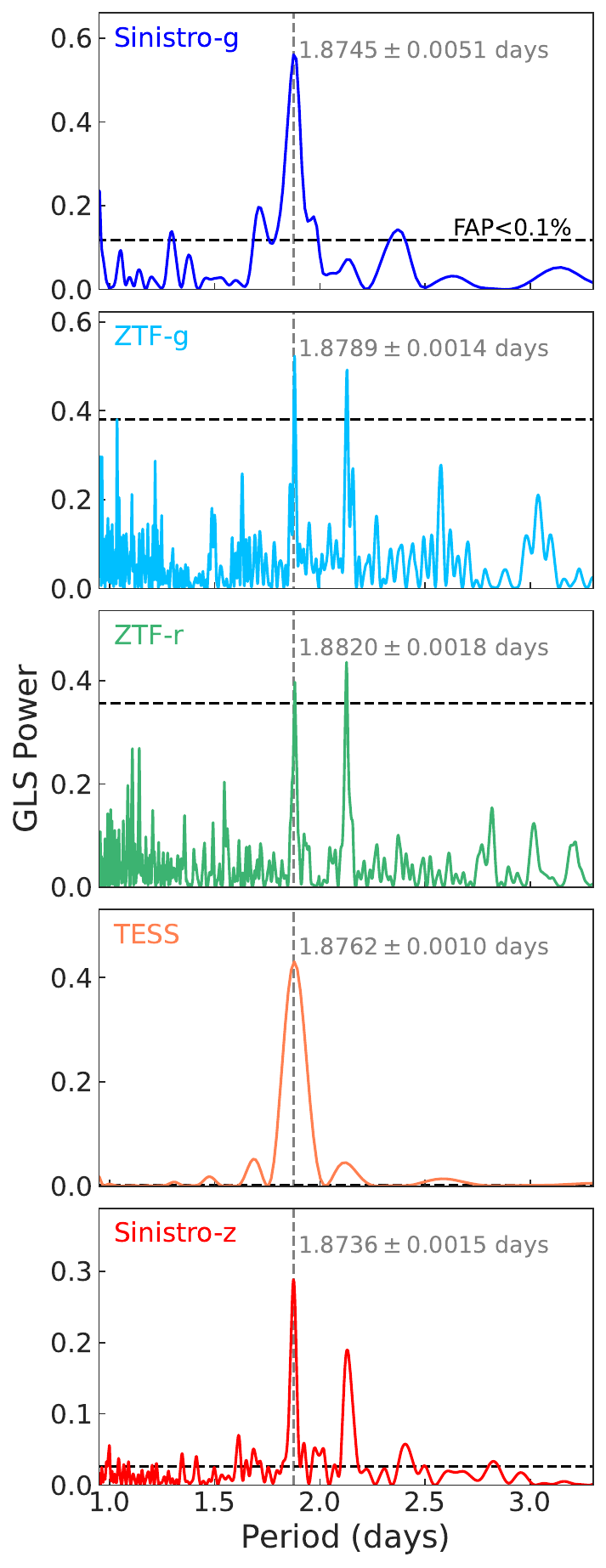}
  \caption{Results of period analysis for Sinistro-$g$, ZTF-$g$, ZTF-$r$, TESS, and Sinistro-$z$ bands. The vertical lines for each panel show the detected period ($1.88$~days). The strong peak at $\sim 2.14$ days show the one-day alias of the signal. The horizontal lines indicate the false alarm probability at 0.1\%. All light curves show clear periodicity.}
  \label{fig:periodogram}
\end{figure}

We derived the periodic signals in the light curves using the Generalized Lomb-Scargle Periodogram \citep[GLS;][]{GLS} in the Python package \texttt{PyPeriod} \citep{pya}. We searched for the period of the light curves between 1 and 20 days and obtained the strong GLS power peak at $\sim 1.88$ days in all light curves (Figure~\ref{fig:periodogram}). False alarm probability (FAP) values below $10^{-3}$ indicate robust detection for those periods. The one-day alias of that signal at $\sim 2.14$ days also shows a strong peak in the power, especially in ZTF light curves, because the ZTF data is collected every few nights.

The mean of the derived periods weighted by the errors was calculated to be $1.87708 \pm 0.00066$~day, and we adopted this value as the stellar rotation period $P_{\rm rot}$. This value is consistent with the previously reported value $P_{\rm rot}=1.88 \pm 0.02$~day from the K2 light curve \citep{Mann2016}. 

\subsection{Monitoring light curve analyses - modulation amplitudes}\label{sec:an-amplitude}

We folded the light curves by the period $P_{\rm rot}=1.87708$ days, assuming that the brightness variability was stable during the observation window and there was no red noise (time-correlated noise) in the light curves in all bands. 
It is possible that instrumental systematics behave as red noise in the light curves, but they are unlikely to be correlated with the stellar rotation phase. The data points were binned every $0.02$ day in all bands in the following analyses. 

We assumed that all the phase-folded light curves have have local maxima and minima at the same phases and differ only in amplitude, and the light curves can be simply modeled as a summation of sine functions. 
The derivation of the amplitudes had two steps. First, we optimized the TESS light curve by a summation of the sine functions as the reference model. We used the TESS light curve here as it has the most clear modulational shape. Second, we modelled the other light curves by changing the amplitude and offset of the reference model.

In the first step, the assumed reference model $f_{\rm TESS}(\bm{p})$ along with the stellar rotational phase $\bm{p}$ is described as
\begin{equation}
    f_{\rm TESS}(\bm{p}) = \sum_{i=1}^{N_s}\left\{a_i \sin 2 \pi( b_i \bm{p} - c_i)\right\},
    \label{eq:sinemodel}
\end{equation}
where $a_i$, $b_i$, and $c_i$ are the parameters to be optimized. The optimization was performed using \texttt{scipy.optimize}, to make the chi-squared value smallest for the TESS light curve. The optimum parameter values are summarized in Table~\ref{tab:sine_constants}. The number of sine-curves $N_s$ was selected to be 2 because the Bayesian information criterion (BIC) became smallest when the $N_s$ value was tested from 1 to 4.

\begin{table}
    \centering
    \begin{tabular}{c|ccc}\hline
               & $a_i$ & $b_i$ & $c_i$\\\hline
         $i=1$ & 0.00892 & 0.5521 & 0.9893 \\
         $i=2$ & 0.00228 &  0.6567 &  1.9693 \\\hline
    \end{tabular}
    \caption{Optimum values for $a_i$, $b_i$, and $c_i$ in Equation~\ref{eq:sinemodel} to fit the phase-folded TESS light curve.}
    \label{tab:sine_constants}
\end{table}

In step two, the flux $\bm{f}_B$ at each band $B=$\{Sinistro-$g$, ZTF-$g$, ZTF-$r$, and Sinistro-$z$\} is represented by the $f_{\rm TESS}(\bm{p})$ with the optimum parameters in Table~\ref{tab:sine_constants}, as in the following equation,
\begin{equation}
    \bm{f}_B (\bm{p}) = A_B \times \frac{f_{\rm TESS}(\bm{p})}{A_{\rm TESS}} + (const.)_B
\end{equation}
where $A_B$ is the amplitude and $(const.)_B$ is the offset at each band. For each band independently, we derived the value that maximises the log-likelihood with two parameters, $A_B$ and $(const.)_B$. Python package \emcee was used for the parameter estimation via the Markov chain Monte Carlo method (MCMC). In this process, we found that the white noise in the light curves are larger than the photometric errors, which may be due to instrumental systematics or observing conditions. In order to correctly account for this noise, we amplified the errors so that the reduced chi-squared value approached $1$, and we re-fit the model to the data with amplified errors.

\begin{figure}
  \centering
      \includegraphics[width=0.4\textwidth]{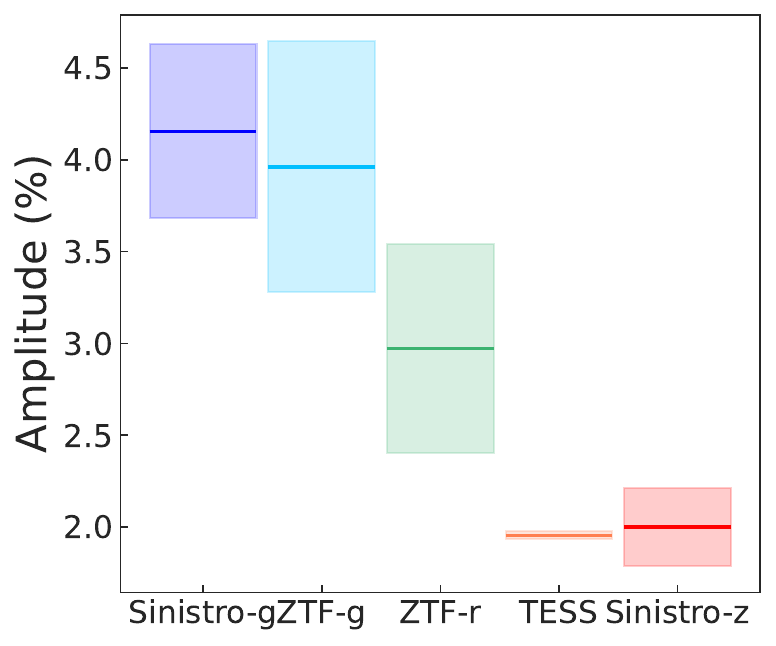}
  \caption{Derived modulation amplitude for the five light curves by sine-curve method (Section~\ref{sec:an-amplitude}). The bands are listed from left to right in order of effective wavelength, from shortest to longest. A clear trend is seen: the longer the wavelength, the smaller the modulation amplitude. The values are summarized in Table~\ref{tab:amplitudes}. }
  \label{fig:amp_sine}
\end{figure}

\begin{figure*}
  \centering
      \includegraphics[width=1.0\textwidth]{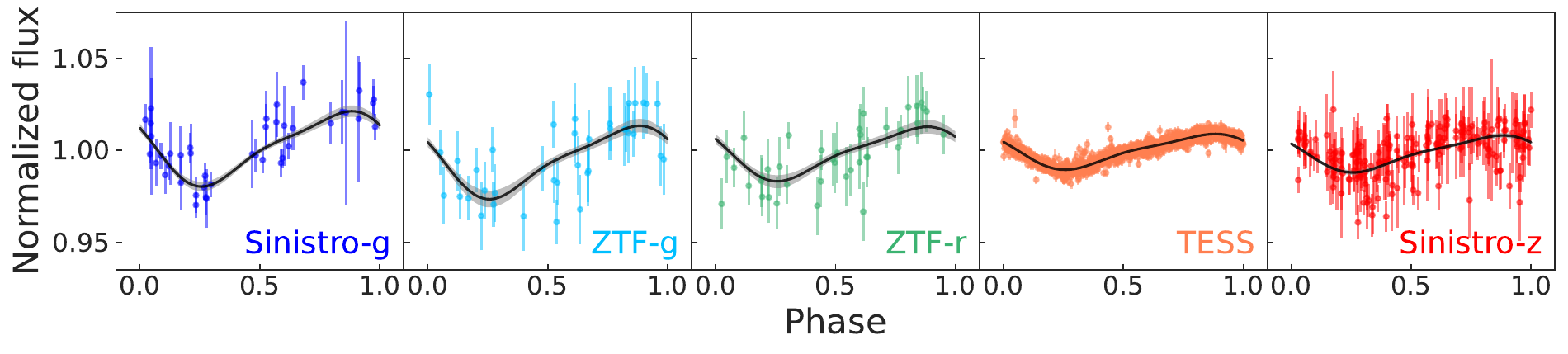}
  \caption{The phase-folded light curves and fitted model curves. The solid line shows the best-fit curves and the shaded region shows a $1 \sigma$ deviation of model uncertainties, derived from the models drawn by the converged MCMC chains.}
  \label{fig:folded-sine}
 \end{figure*}

Figure~\ref{fig:amp_sine} summarizes the resulting $A_B$ values with their uncertainties, and Figure~\ref{fig:folded-sine} shows the phase-folded light curves plotted with the derived best-fit models with 1$\sigma$ uncertainties. It is clearly seen that the amplitude becomes smaller to longer wavelengths. As the Sinistro-$g$ and ZTF-$g$ have almost identical wavelength ranges, the weighted-mean value is calculated and used in the following analyses. We also tried another method using a Gaussian process (GP) model, which is more flexible for modelling the light curves, to independently derive the modulation amplitudes (see Appendix~\ref{sec:ap-gp}) and confirmed that the results are consistent. 

\subsection{Derivation of spot temperature and covering fraction}\label{sec:an-spottemp}
The differences in the amplitude at each wavelength reflect differences in the spot contrast, defined as the spot intensity relative to the photosphere. Based on that, we derive the spot temperature and covering fraction from the modulation amplitudes in this section.

We note that spots visible regardless of the stellar rotation phase (referred to as ``always-visible spots" hereafter) are not considered in this analysis. It has been noted that even with multi-band light curves, constraining the covering fraction of always-visible spots is challenging \citep{Rosich2020}. We also confirmed that the estimation of the spot temperature and spot covering fraction, which contributes to the brightness variation, remains unaffected whether the presence of always-visible spots is considered or not. On the other hand, while always-visible spots do not contribute to the brightness variation, they do cause TLSE during planetary transits. We account for them in the TLSE estimation in Section~\ref{sec:dis-tlse-stack} by assuming various values of spot covering fraction. 

\begin{figure}
\centering
  \includegraphics[width=0.4\textwidth]{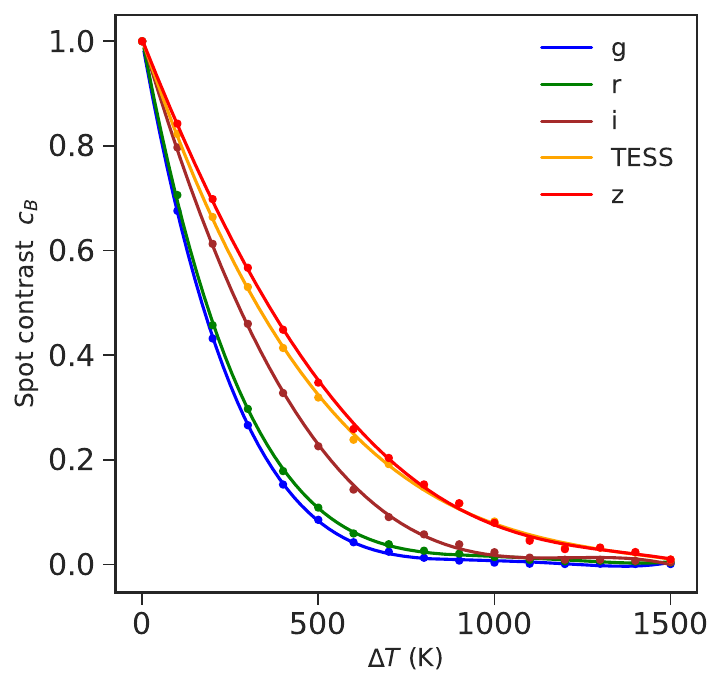}
\caption{Relation of calculated spot contrast $c_B$ to $\Delta T$ for each observed band $B=\{g, r, i, z, \rm{TESS}\}$. $\Delta T$ represents the difference of spot temperature from the stellar surface temperature fixed at 3200\,K. The wavelength dependence of the value of $c_B$ is greatest around $\Delta T=500$\,K. }
\label{fig:contrast}
\end{figure}

Assuming that spots are scarcely visible during the peak brightness of a star and the spots with temperature $T_{\rm spot}$ and covering fraction $f_{\rm spot}$ during the dimmest phase, the amplitude of stellar brightness variation $A_B$ at each band $B= \{g, r, {\rm TESS}, z\}$ can be expressed by
\begin{equation}
    A_B = |1-c_B| f_{\rm spot} \label{eq:fs}
\end{equation}
where $c_B$ are the spot contrast \citep[e.g., Equation 8 in][]{Notsu2013}.
The spot contrast $c_B$ is calculated from the following formula \citep[e.g.,][]{Morris2018,Ikuta2023}:
\begin{equation} 
c_B = \frac{\int_{\rm B} d\lambda \mathcal{T}_\lambda \mathcal{F}_{\lambda, \rm{spot}}} {\int_{\rm B} d\lambda \mathcal{T}_\lambda \mathcal{F}_{\lambda, \rm{phot}}}, \label{eq:morris}
\end{equation}
where $\mathcal{T}_\lambda$ is the filter transmission function and $\mathcal{F}_{\lambda, \rm{spot/phot}}$ are the stellar flux from the spots or the photosphere at the temperature of $T_{\rm spot}$ and $T_{\rm phot}$, respectively. We calculated $c_B$ through Equation~\ref{eq:morris} using the filter profiles from the SVO Filter Profile Service \citep{SVO1, SVO2} and theoretical stellar spectra from the SVO Theoretical Model Services. For stellar spectra models, we used the BT-Settl model with the surface gravity $\log g = 5$ (in cgs), and the metallicity [Fe/H] $=$ 0 dex. The spot temperature $T_{\rm spot}$ ranged from 2700\,K to 3200\,K in each 100\,K step for the calculation and then interpolated. The photosphere temperature $T_{\rm phot}$ was fixed to be 3200\,K, but we confirmed that the $\sim 100$\,K difference of $T_{\rm phot}$ did not cause significant change to the relation of $T_{\rm spot}$ and $c_B$. Figure~\ref{fig:contrast} shows the relation of calculated $c_B$ and $\Delta T$, which is defined by $T_{\rm phot} - T_{\rm spot}$, for all observed bands. The $c_B $values show band-dependent variation when $\Delta T$ is around 500\,K, but the variation gets smaller when $\Delta T$ is close to 0\,K or more than 1000\,K, in the case of $T_{\rm phot}=3200$\,K. 

The combination of Equations~\ref{eq:fs} and \ref{eq:morris} enables us to estimate the two unknown values, $T_{\rm spot}$ and $f_{\rm spot}$ from the estimated amplitude $A_B$ in each band. As there are four bands in our analysis, $T_{\rm spot}$ and $f_{\rm spot}$ can be obtained simultaneously by formulating four of these equations with each error. The parameters are set to $\Delta T$ and $f_{\rm spot}$, and the best parameter values were searched to describe the observed amplitudes for the four bands by calculating chi-squared values. We used MCMC package \emcee for the parameter estimation. We put prior for $\Delta T$ to take positive values and $f_{\rm spot}$ to take values between 0 to 1.

\begin{figure}
\centering
      \includegraphics[width=0.47\textwidth]{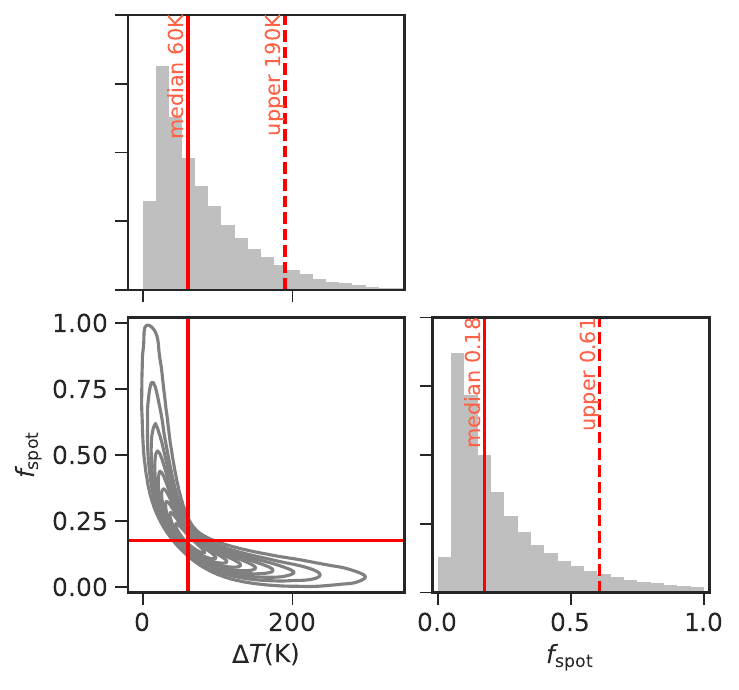}
  \caption{Posterior distribution of the spot temperature difference $\Delta T = T_{\rm phot} - T_{\rm spot}$ and the spot covering fraction $f_{\rm spot}$. The red solid lines show the median values and the dashed lines show the upper limit with the 2$\sigma$ level.}
  \label{fig:corner-simple}
 \end{figure}

Figure~\ref{fig:corner-simple} shows the posterior distribution of the parameters using \corner \citep{corner}. Although they are highly correlated, we derived the $2 \sigma$ upper limit of $\Delta T$ is \upperT\,K and $f_{\rm spot}$ is \upperf\%. Note that we did not consider the possibility of spots with temperature higher than the stellar photosphere (referred to as ``bright spots"). We discuss the case for bright spots in Appendix~\ref{ap:bright}.

\subsection{Investigation of the spot distribution}\label{sec:an-distribution}
To evaluate the analysis of estimating spot temperature and covering fraction only from the amplitudes in multi-band light curves (Section~\ref{sec:an-spottemp}), we further explored possible spot distributions using a spot mapping method \citep[][]{Ikuta2020,Ikuta2023}. Given that there are known degeneracy in various spot distributions as a solution to the starspot mapping from the light curves \citep{Basri2020,Ikuta2020,Luger2021_mapping1}, this study adopted the method of trying out several patterns of spot distribution and exploring their trends rather than searching for a single solution. 

We used an analytical spotted model \texttt{macula} \citep{Kipping2012} as in \citet{Ikuta2023}. 
It calculates the flux relative to the spotless photosphere assuming a particular stellar inclination angle, for a specified number of circular spots characterized by four parameters: spot contrast, radius, latitude, and longitude \citep[for more details, see][]{Ikuta2023}. 

We calculated the light curves in $g$, $r$, TESS, and $z$-band under different spot contrast and stellar limb-darkening coefficients. 
The spot temperature was set to be identical for all spots. The spot contrast at each band was calculated by Equation~\ref{eq:morris} under the assumed temperature.
The limb-darkening coefficients were calculated by the package \ldtk \citep{Parviainen2015_ldtk,Husser2013}, assuming the stellar effective temperature $T_{\rm eff}=3207\pm 58 \,{\rm K}$, metallicity ${\rm [Fe/H]}=0.15\pm 0.03 \,{\rm dex}$, and stellar surface gravity $\log g = 4.944\pm0.031$ (in cgs).
The stellar inclination and the degree of differential rotation were set to 90 deg \citep{Stefansson2020} and 0. The rotation period was set to 1 because the light curves were phase-folded.

We reproduced various patterns of stellar surface models, with the number of spots set to two, three, four, and five. In each spot configuration with the number of spots of $N_s$, the light curve model was generated with $3 N_s+1$ parameters: radius, latitude, and longitude for each spot and $\Delta T$. Since a larger number of parameters require a higher computational cost to obtain the optimal values, we provided some additional settings to simplify the calculations. First, we fixed the $T_{\rm phot}$ at 3200\,K. Also, the latitude of any spot was restricted to taking only positive values. It is because the stellar inclination of K2-25 is $\sim$ 90 deg so it is impossible to determine whether a spot is on the northern or southern hemisphere of the star. For spot size, we set uniform prior so that the maximum spot radius was 30$^\circ$. We limited $\Delta T$ to be a positive value so that faculae (or any bright spots) were not considered.

At each number of spots $N_s$, we derived the posterior distributions using \emcee to fit the reproduced light curves with the observed ones by calculating the chi-squared values. In the calculation, we adopted the amplified uncertainties of the observed light curves calculated in Section~\ref{sec:an-amplitude}. 

In addition, to compare the results by spot mapping with those from the amplitude, we calculated the spot covering fraction $f_{\rm spot}$ from the posterior distribution, by the summation of the projected area of each spot \citep[Equation 11 in][]{Ikuta2020} at the minimum of the light curve.

\begin{figure} 
\centering
  \includegraphics[width=0.47\textwidth]{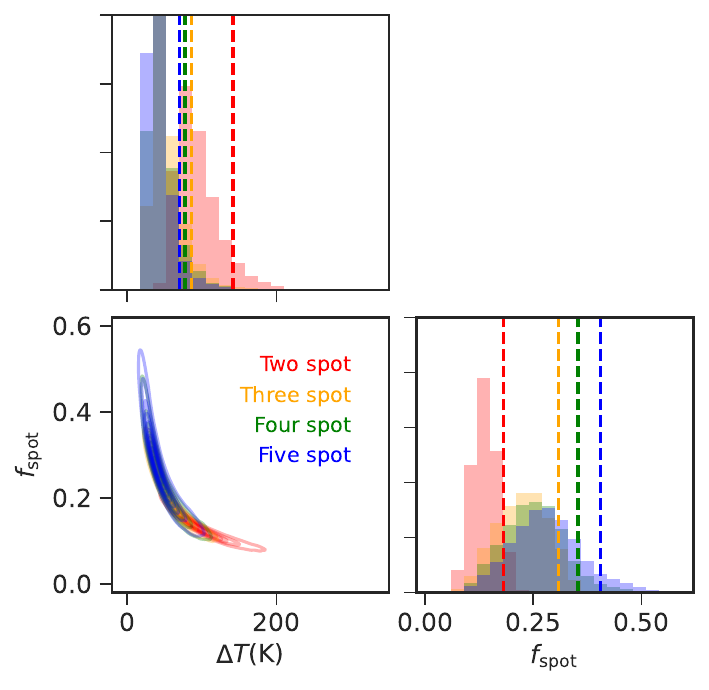}
\caption{Posterior distribution of $\Delta T$ and $f_{\rm spot}$, with two, three, four, and five-spot cases shown in red, orange, green and blue. The dashed lines indicate the $2\sigma$ upper limit of $\Delta T$ and $f_{\rm spot}$ at each case.}
\label{fig:corner_spotnumber}
\end{figure}

Figure~\ref{fig:corner_spotnumber} shows the posterior distributions of $\Delta T$ and $f_{\rm spot}$ at each spot number case using \corner \citep{corner}. The $2\sigma$ upper limit of $\Delta T$ was $142, 86, 78, 70$\,K and the upper limit of $f_{\rm spot}$ was $18, 31, 35, 41\%$ for two, three, four and five spots cases, respectively. 

While the two-spot case showed slightly different values, the upper limit values were similar for more than three spots. The model evidence was calculated via the importance sampling algorithm \citep[e.g.,][]{Diaz2014} as well as \cite{Ikuta2020}. 
The logarithm of the model evidence was calculated to be $-711.34$, $-707.81$, $-706.60$, and $-705.78$ for two, three, four, and five-spot models, respectively.
The model evidence is almost equal for the three, four, and five-spot cases within a factor of 8 ($\sim \exp(2.02)$), while the two-spot case showed relatively lower model evidence by three orders of magnitude ($\sim \exp(5.56)$) than the five-spot case.
Then, the model with more than three spots is preferred in the model comparison \citep{Kass1995}. The joint posterior for the three-spot case is shown in Figure~\ref{fig:corners_all} using \corner \citep{corner}.

\begin{figure*}
\begin{minipage}[b]{0.45\linewidth}
    \centering
    \includegraphics[width=1\textwidth]{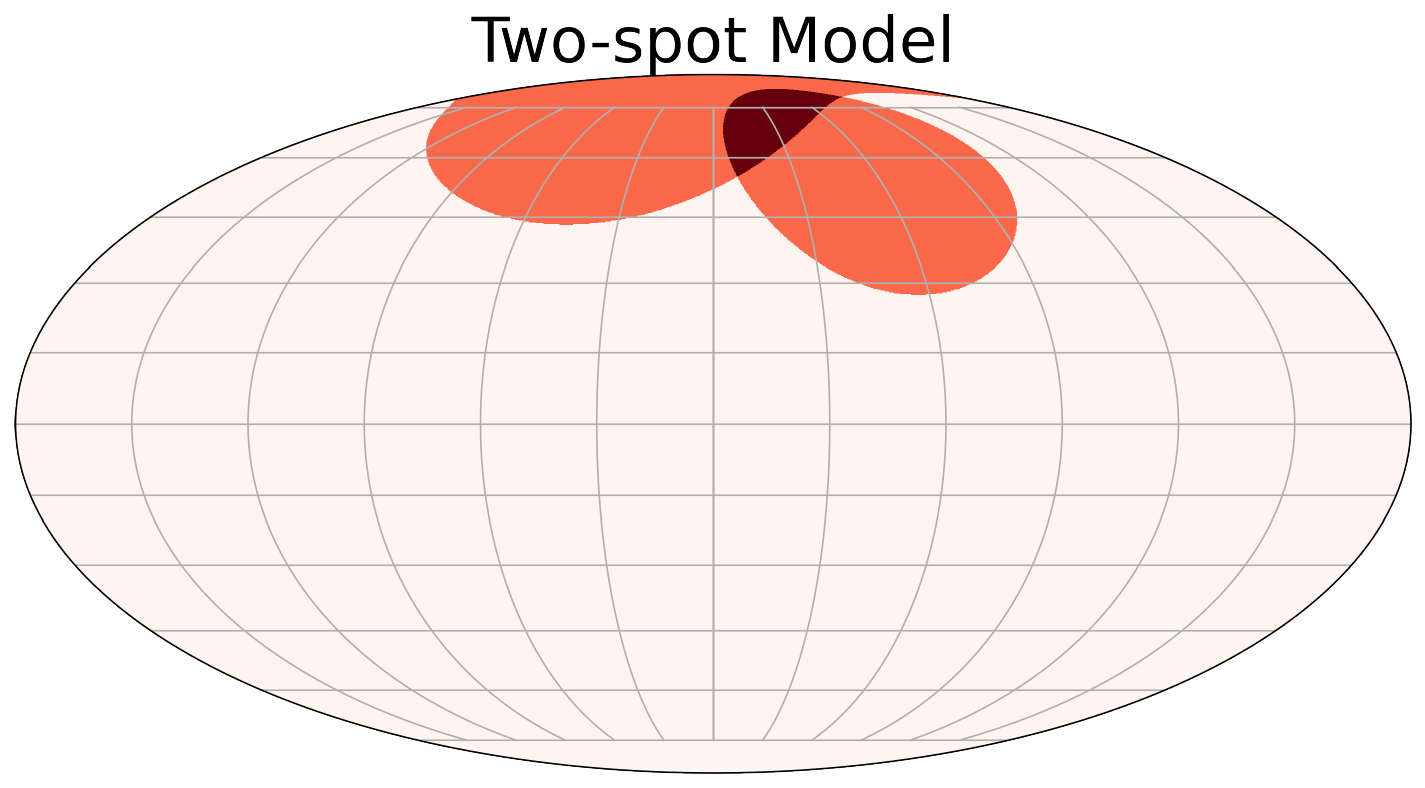}
  \end{minipage}
  \begin{minipage}[b]{0.45\linewidth}
    \centering
    \includegraphics[width=1\textwidth]{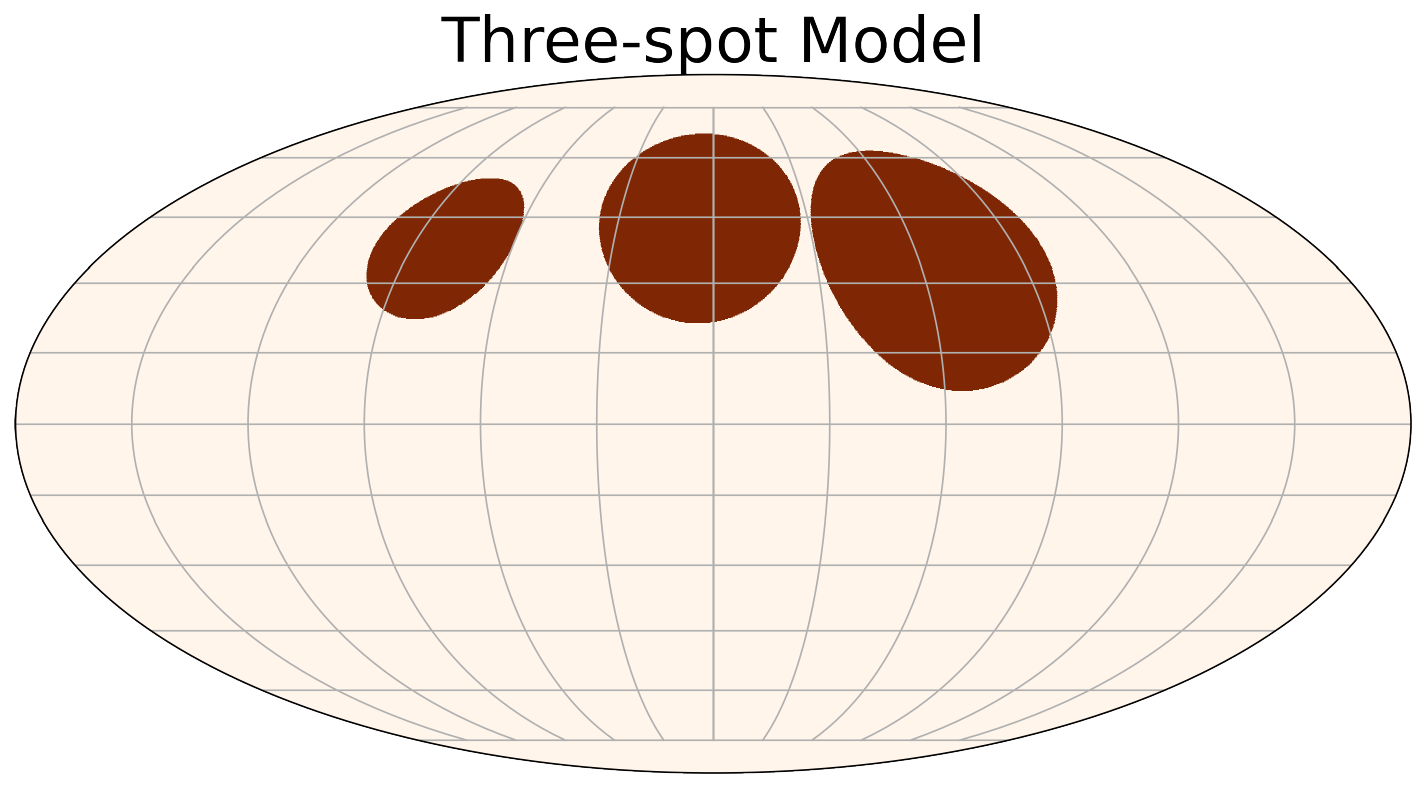}
\end{minipage}
\begin{minipage}[b]{0.45\linewidth}
    \centering
    \includegraphics[width=1\textwidth]{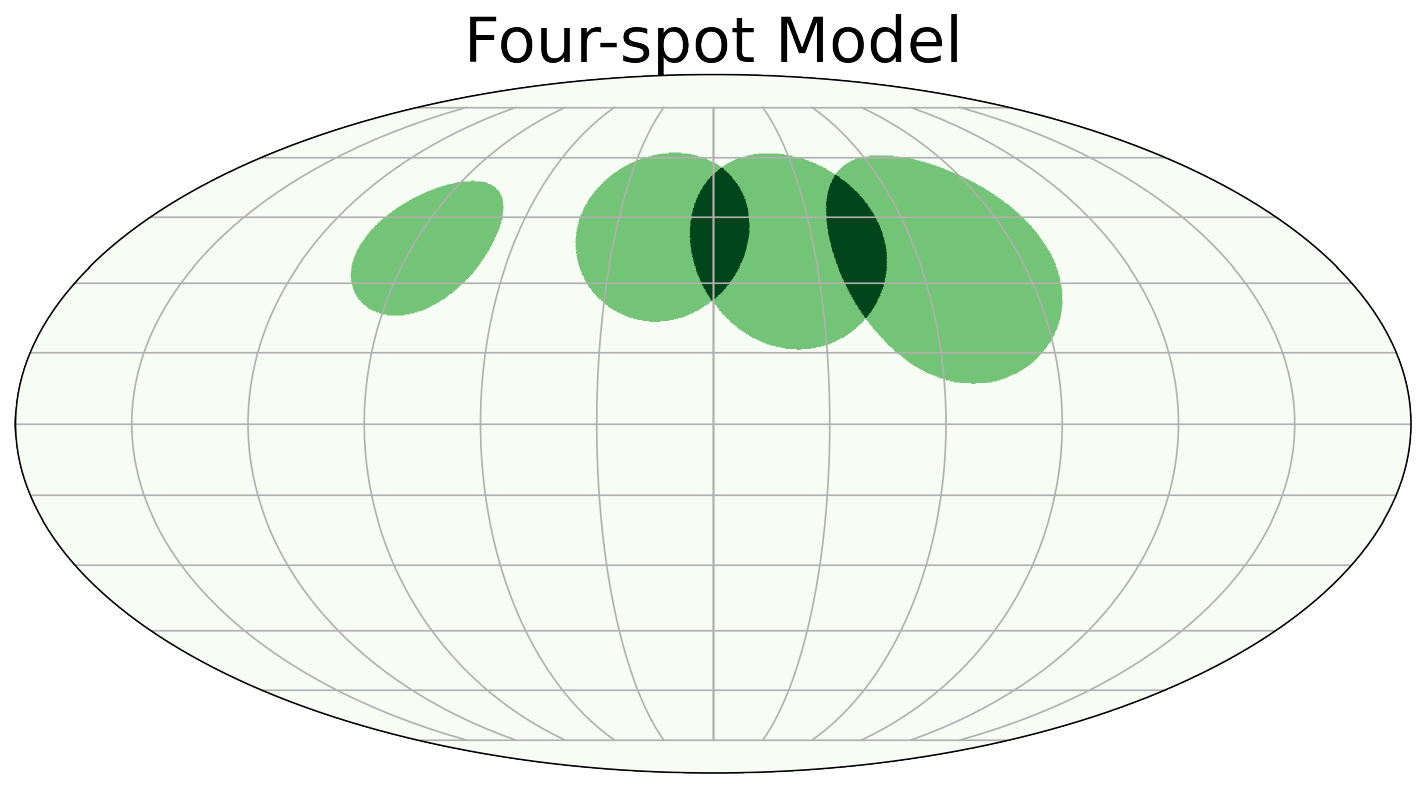}
  \end{minipage}
  \begin{minipage}[b]{0.45\linewidth}
    \centering
    \includegraphics[width=1\textwidth]{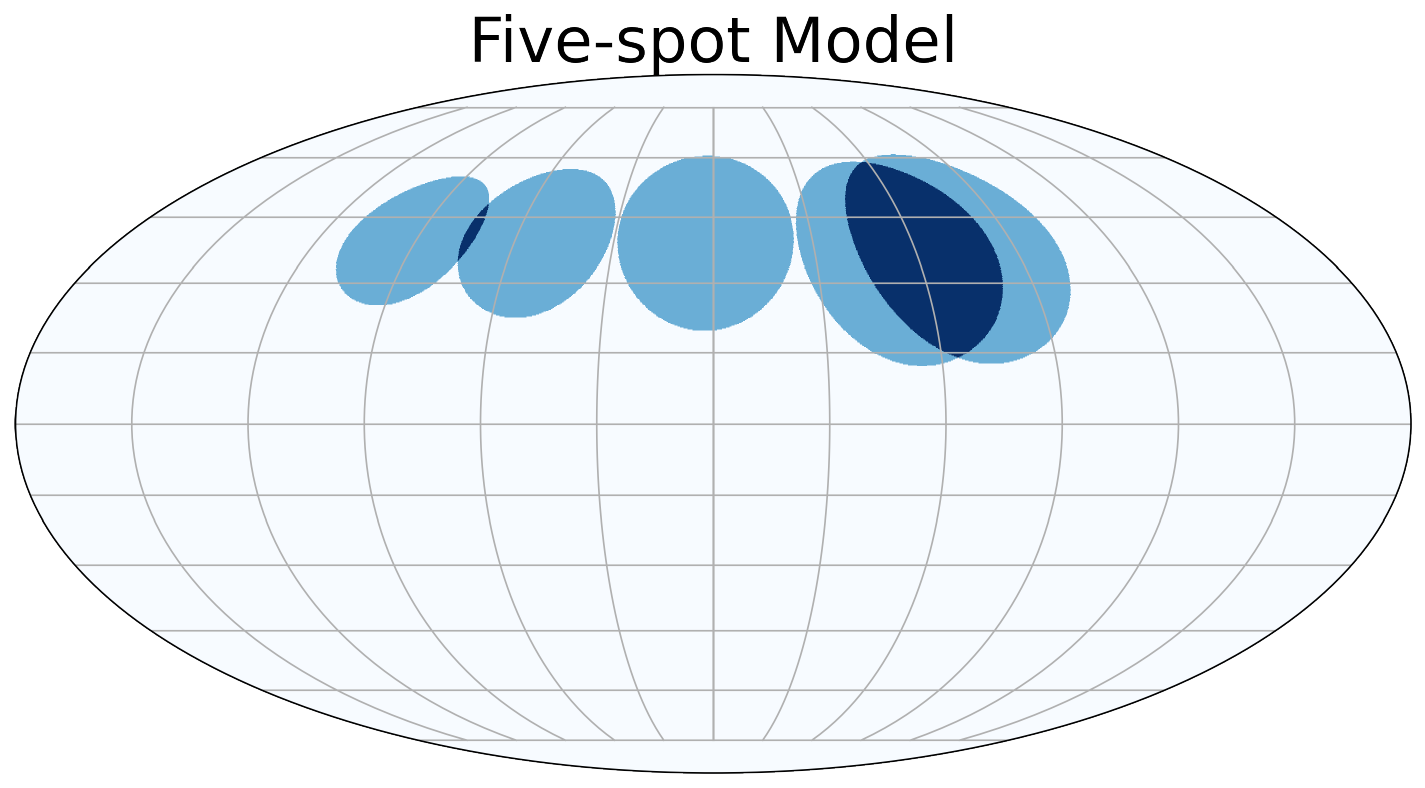}
\end{minipage}
\caption{Spot map with optimum spot parameters for the two, three, four, and five-spot cases shown in red, orange, green and blue, respectively. The spots are located at high altitudes in the two-spot case. In cases with a larger number of spots, the spots show a similar distribution in the mid-latitude zone.}
\label{fig:spotmaps}
\end{figure*}

Figure~\ref{fig:spotmaps} visualises the best-fit spot distribution at phase = 0 in the Mollweide projection for the two, three, four and five-spot cases. In the case of two spots, spots appear in the higher latitude ($>60$ deg for both spots in the best-fit values) whereas in the case of three or more spots, they are found to be distributed similarly at about similar latitudes around $38$ deg, while each spot latitude value has large ($>15$ deg) uncertainties. In any case, the optimal solution preferred a pattern where the spot is barely visible when the star is brightest.

We note that our calculation does not take spot overlaps into account (i.e. the overlapping spots are calculated to be twice as dark). The spot latitude is restricted to take only positive values (northern hemisphere) because of the degeneracy with negative values (southern hemisphere) under the stellar inclination of $\sim$ 90 deg.
This overlap could be eliminated if one of the overlapped spots is placed alternately on the southern hemispheres. 

The resulting upper limit of spot temperature and spot covering fraction (Figure~\ref{fig:corner_spotnumber}) by spot mapping was consistent with the result of amplitude-spot temperature relations (Figure~\ref{fig:corner-simple}). We will further discuss the results in Section~\ref{sec:dis-spots}.

\subsection{Estimation of transit light source effect} \label{sec:dis-tlse-stack}

From the derived spot characteristics, we estimated how much the observed transit depths were affected by spots.
The degree of TLSE to the transit depth at each wavelength is quantified by the contamination factor $\epsilon_\lambda$ \citep{Rackham2017} as follows:
\begin{equation}
    \delta_{\lambda, \rm obs} = \epsilon_\lambda \delta_{\lambda, \rm true},
\end{equation}
where $\delta_{\lambda, \rm obs}$ is the observed transit depth at each wavelength and $\delta_{\lambda, \rm true}$ is the true transit depth originating from the planetary atmosphere. From \cite{Rackham2018}, the $\epsilon_\lambda$ value is determined by the ratio of the flux from the photosphere to the flux from the transit chord. Simply, it is calculated by the equation
\begin{equation}
\epsilon_\lambda = \frac{\mathcal{F}_{\lambda, \rm{phot}}}{f_{\rm spot}\mathcal{F}_{\lambda, \rm{spot}}+(1-f_{\rm spot})\mathcal{F}_{\lambda, \rm{phot}}} . \label{eq:tlse-contrast}
\end{equation}
For each band, the equation can be represented as
\begin{equation}
\epsilon_B = \frac{1}{1-f_{\rm spot}(1-c_B)} . \label{eq:tlse-contrast-band}
\end{equation}
using the spot covering fraction $f_{\rm spot}$ and spot contrast $c_B$ at the band $B$, obtained from equations~(\ref{eq:fs}) and (\ref{eq:morris}) (Section~\ref{sec:an-spottemp}). Here, we assume that there are no spots in the transit chord. This should be a reasonable assumption based on the fact that our observations show no signs of spot crossings (see Appendix~\ref{sec:an-crossing}).

Figure~\ref{fig:varepsilon} shows the contamination factor (Equation~\ref{eq:tlse-contrast}) for each wavelength, assuming spots with a temperature of 3100\,K on the stellar surface with a temperature of 3200\,K as a reasonable assumption from our estimation of $\Delta T$ from the stellar brightness variability (Sections~\ref{sec:an-spottemp} and \ref{sec:an-distribution}). As the $f_{\rm spot}$ can be larger if there are always-visible spots, we calculated the TLSE with different spot covering fractions from 5 \% to 30 \%. It is clearly seen that TLSE is larger in the optical wavelength range than in the infrared. In the near-infrared wavelength range that is important for the atmosphere characterization, the value of $\epsilon_\lambda$ varies by up to 1\%; as the transit depth of K2-25 is $\sim 1$\%, this is expected to result in a transit depth variation of 100 ppm.

Separately, TLSE can also be calculated from the mapping results. Since Equation~\ref{eq:tlse-contrast} represents the ratio of the flux coming from the spotless surface to the current stellar surface, we can divide the maximum value of the flux by the flux coming from the star at each rotation phase derived from the spot mapping. In this approach, the information regarding the spot positions and stellar limb darkening, which are known to potentially influence the estimation of TLSE\citep{Thompson2024}, was incorporated, as it directly utilizes the results of spot mapping. We confirmed that the typical TLSE value derived by this method was comparable to the value calculated by the method described above. This approach also enables the prediction of TLSE variations across different stellar rotational phases. Further details are given in the Appendix~\ref{sec:ap-depvar}.

\begin{figure*}
  \centering
      \includegraphics[width=0.9\textwidth]{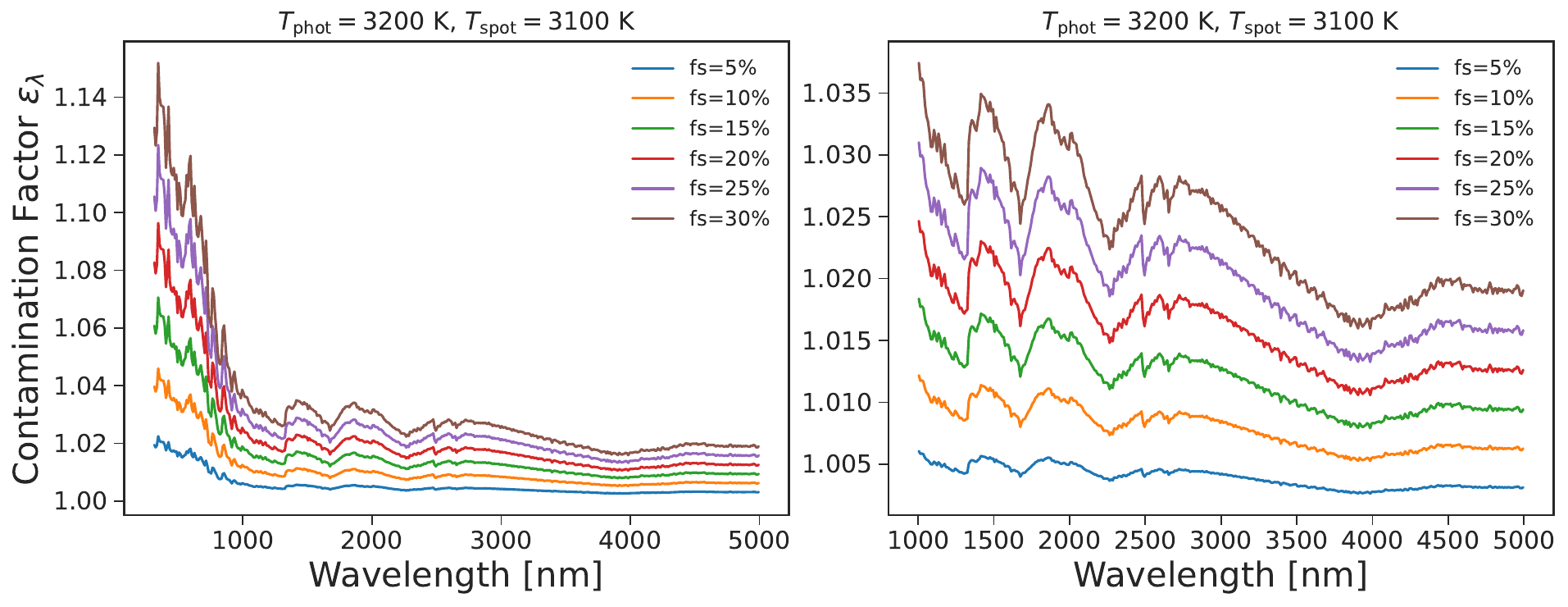}
  \caption{Calculated TLSE for each possible spot covering fraction $f_{\rm spot}$, assuming the spots with $T_{\rm spot}=3100$\,K on the photosphere temperature $T_{\rm phot} =3200$\,K. The right figure is just a close-in to the wavelength over 1000nm. TLSE works strongly in the optical, and it fluctuates the transit depth by about 1\% in the near-infrared.}
  \label{fig:varepsilon}
\end{figure*}

\subsection{Transit light curve analyses}\label{sec:an-transit}
\begin{figure*}
  \centering
      \includegraphics[width=1.0\textwidth]{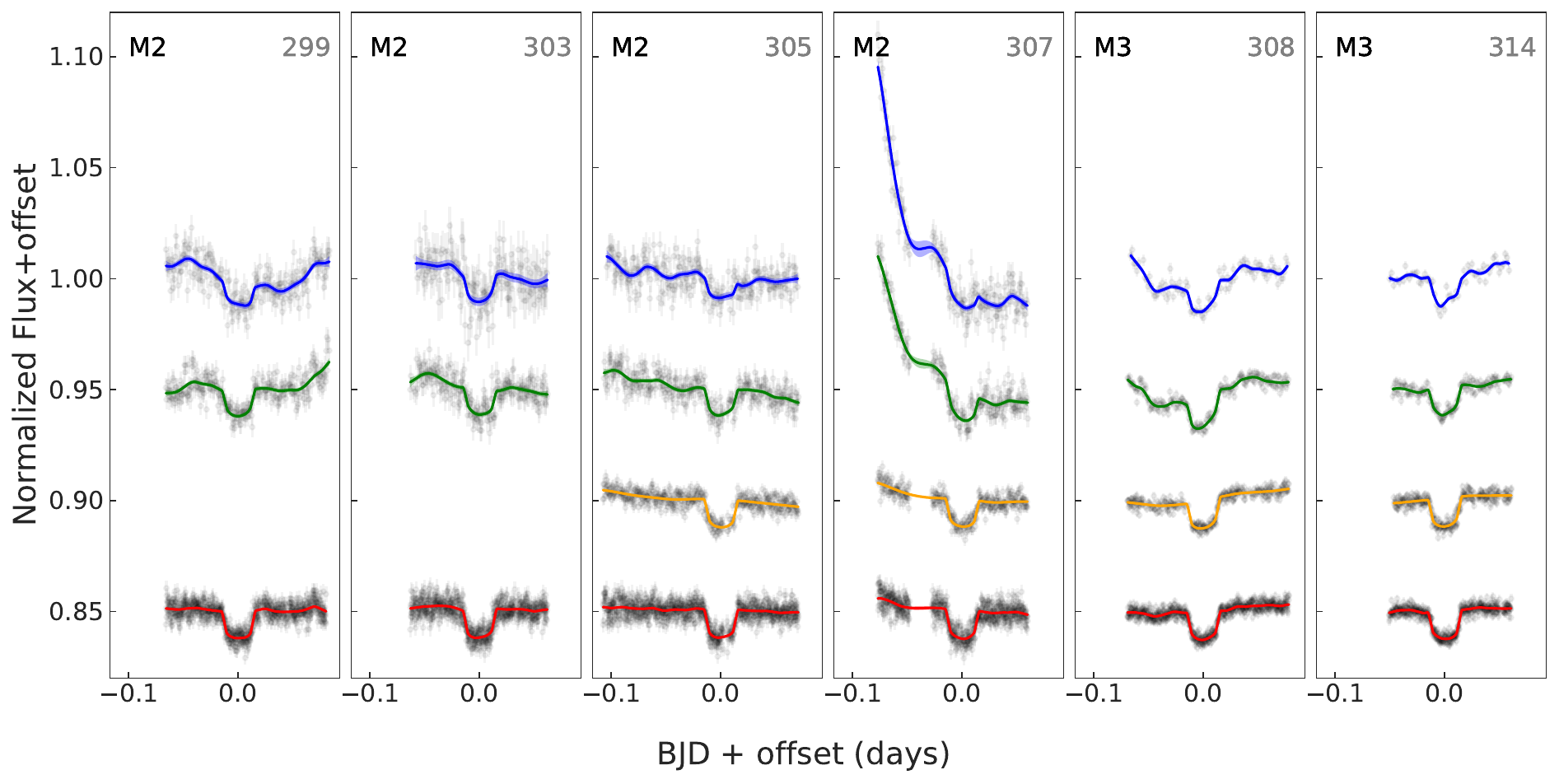}
  \caption{Transit light curves of K2-25\,b taken with MuSCAT2 \& 3 in $g$, $r$, $i$, $z$-band (from top to bottom) on each observing night. $N_{\rm T}$ number is shown next to each light curve. In each subplot, observed data points are shown in grey points, and best-fit transit + baseline models derived by GP regression are indicated in coloured lines. The shaded regions indicate the 1$\sigma$ error of the GP model.}
  \label{fig:whole_transit}
 \end{figure*}

In this section, we focus on the transit light curves taken by the MuSCAT2, MuSCAT3, and TESS (summarized in Table \ref{tab:observations}). MuSCAT2 and MuSCAT3 observed the transits of K2-25\,b six times in $g$, $r$, and $z$ band, and four times in $i$ band. TESS observed the six transits. Therefore, we have 28 transit light curves in total (Figure~\ref{fig:whole_transit}). We analysed these transit light curves to derive the transit depths at each band to evaluate the degree of TLSE.

We jointly fit all 28 light curves using the mean models defined by the transit parameters and noise models defined as Gaussian processes (GP). The parameters, priors, and derived values from the following procedures are summarized in Table~\ref{tab:trfit-results-eccentric}. 

For the mean model of the planetary transit, we used the package \batman \citep{BATMAN}, which is based on the analytic formula in \citet{MandelAgol}. As transit parameters, we adopt the transit-centre time $T_0$, the orbital period $P_{\rm orb}$, the orbital semi-major axis over the stellar radius $a/R_\star$, the impact parameter $b$, and the ratio of the planetary radius to the stellar radius $R_p/R_\star$. Only $R_p/R_\star$ values were set as different parameters for each band. We assumed a linear ephemeris, as we did not detect any transit timing variations from our light curves. 
As the previous studies have shown K2-25\,b should have an eccentric orbit \citep{Thao2020, Stefansson2020}, we added two more parameters $\sqrt{e} \cos \omega$ and $\sqrt{e} \sin \omega$, where $e$ and $\omega$ are the eccentricity and longitude of periastron, respectively. In addition, modified quadratic limb-darkening coefficients $(q_1, q_2)$ were set as parameters for each band \citep{Kipping2013}. We used normal priors (Gaussian priors) for $T_0$, $P$, $b$ based on the values in \citet{Stefansson2020}, and normal priors for the limb-darkening coefficients based on the theoretical calculation by the package \ldtk assuming the stellar effective temperature $T_{\rm eff}=3207\pm 58 \,{\rm K}$, metallicity ${\rm [Fe/H]}=0.15\pm 0.03 \,{\rm dex}$, and stellar surface gravity $\log g = 4.944\pm0.031$ (in cgs). In addition, we set the normal prior for stellar density $\rho_\star$ to the value derived from \citet{Stefansson2020}. Uniform priors were set for the other parameters.

To account for baseline trends in light curves, we used the GP with an approximated Mat\'ern-3/2 kernel:
\begin{equation}
    k_{MA}(\tau)= \frac{1}{2}\sigma^2 \left[(1+1/\varepsilon)e^{-(1-\varepsilon)\sqrt{3}\tau/\rho}+(1-1/\varepsilon)e^{-(1+\varepsilon)\sqrt{3}\tau/\rho}\right]
\end{equation}
which is implemented in \celerite \citep{celerite1,celerite2}. Here, $\tau$ is the absolute time difference of each observing point, meaning that a data point is correlated with another data point. The hyperparameters $\sigma$ and $\rho$ determine the correlation of the data points: $\sigma$ corresponds to the amplitude of correlation strength, and $\rho$ is the timescale of the exponential decay. $\varepsilon$ was fixed to 0.01 by default. Assuming that the baseline variation is caused by the brightness variation of the star itself or the wavelength-dependent instrument noise, we set hyperparameters for each band. As we do not know any prior information on these hyperparameter values, we set log-uniform priors for the hyperparameters.

\begin{figure*}
  \centering      \includegraphics[width=1.0\textwidth]{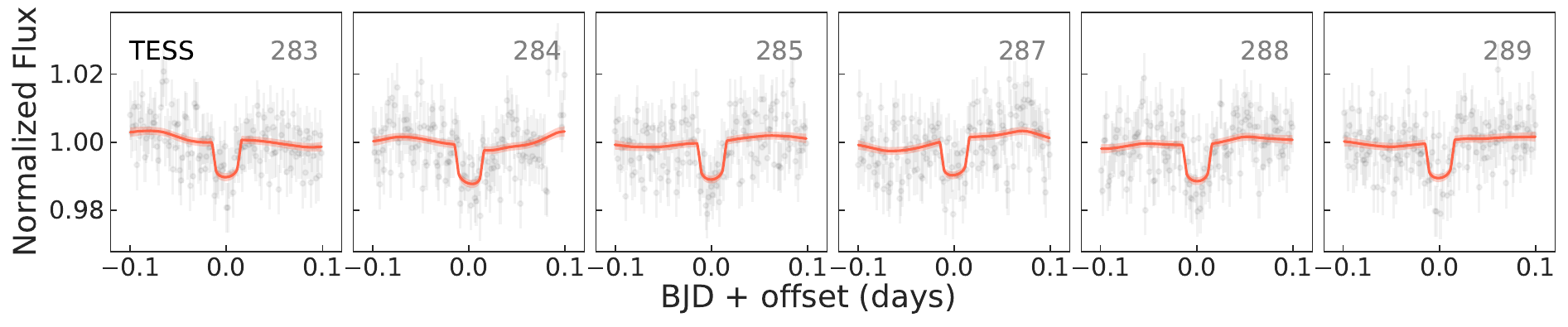}
  \caption{Similar to Figure~\ref{fig:whole_transit} but for transit light curves taken with TESS. Note that TESS observations are not simultaneous with the MuSCAT2/3 observations.}
  \label{fig:whole_transit_TESS}
 \end{figure*}

As shown in Figures~\ref{fig:whole_transit} and \ref{fig:whole_transit_TESS}, we jointly optimized 28 transit light curves in total with the mean and noise models. The total number of parameters and hyperparameters were summed up to 31. The parameter estimation was performed using \emcee, and the derived values are listed in Table~\ref{tab:trfit-results-eccentric}. The derived transit depths for each band are also shown in Figure~\ref{fig:platon}. Considering the large uncertainties in $g$ and TESS-band, we do not see significant transit depth dependence on the colour.

\subsection{Constraints on spot temperature and covering fraction from transit depths}\label{sec:an-tlsetransit}
As shown in Figure~\ref{fig:varepsilon}, the TLSE by dark spots makes the observed transit depths larger in shorter-wavelength bands in the optical. The derived transit depths did not show such a trend, although their uncertainties are large. Independently of the method obtained from the monitoring light curves, we investigated the possible spot characteristics only from the transit depths observed by MuSCAT2/3 and TESS.

Assuming the true transit depth is constant for all observed bands (i.e., flat transmission spectra model), 
The observed transit depths at each band $\delta_B$ can be described as 
\begin{equation}
    \delta_B = \epsilon_B \delta_{\rm true},
\end{equation}
where $\epsilon_B$ is a contamination factor for each band and $\delta_{\rm true}$ is a true transit depth (constant value). 
$\epsilon_B$ for each band is calculated through the Equation~\ref{eq:tlse-contrast-band}. We modelled the $\delta_B$ for band $B=\{g, r, i, z, \rm{TESS}\}$ with three parameters, $\Delta T$, $f_{\rm spot}$, and $\delta_{\rm true}$. We set uniform prior $\mathcal{U}(0, 1500)$ for $\Delta T$ and $\mathcal{U}(0, 1)$ for $f_{\rm spot}$. $\delta_{\rm true}$ was restricted to take only positive values. The parameter estimation was done using \emcee.

\begin{figure}
  \centering
      \includegraphics[width=0.48\textwidth]{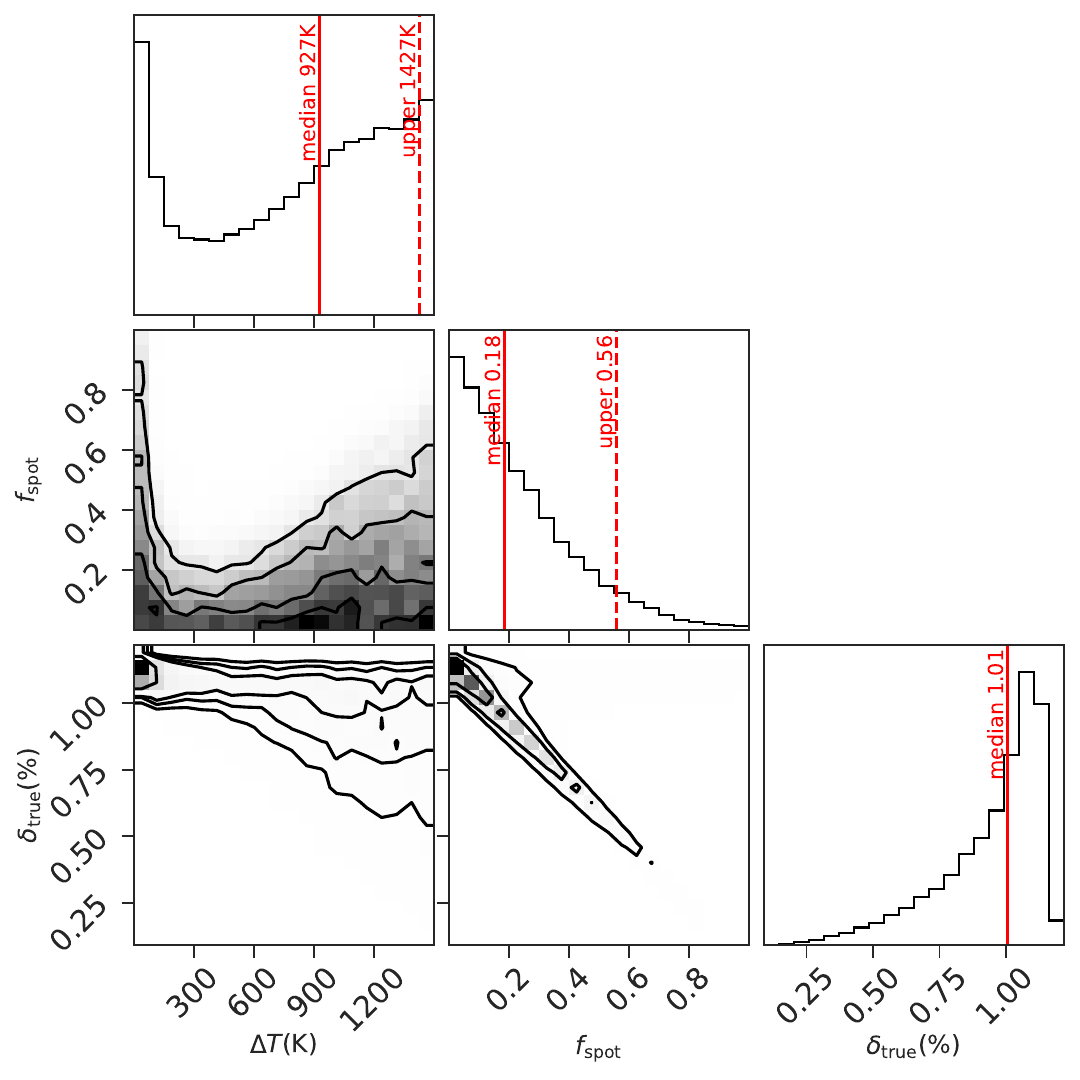}
  \caption{Posterior distribution of the spot temperature difference $\Delta T$, the spot covering fraction $f_{\rm spot}$, and the constant transit depth to explain the observed transit depths.}
  \label{fig:corner_transit}
\end{figure}

Figure~\ref{fig:corner_transit} shows the derived posterior distribution using \corner \citep{corner}. The $2\sigma$ upper limit of $\Delta T$ and $f_{\rm spot}$ values were 1427\,K and 0.56, respectively. 
This means that little restriction could be placed on the spot characteristics from only the wavelength dependence of the transit depths. The posterior of $\Delta T$ shows a weak bimodal distribution, which is attributed to the fact that the obtained transit depths are consistent with flat considering the uncertainties and does not show a strong wavelength dependence. This is because the wavelength dependence of the spot contrast values is small at both low and high $\Delta T$ values, as shown in Figure~\ref{fig:contrast}. However, for large values of $\Delta T$, either very small $f_{\rm spot}$ values or $\delta_{\rm true}$ values will be required, which is unlikely in light of the other observed facts. Nevertheless, models with low-temperature spots spread over a wide area were ruled out. These results were consistent with the results from the monitoring light curves (Figure~\ref{fig:corner-simple}, \ref{fig:corner_spotnumber}).

We also tried to derive $\Delta T$, $f_{\rm spot}$, and $\delta_{\rm true}$ by jointly fitting the transit depths and modulation amplitudes (Section~\ref{sec:an-spottemp}). The derived posterior distribution of the parameters was almost identical to the results from the fitting of modulation amplitudes (Figure~\ref{fig:corner-simple}), meaning that the spot characteristics are constrained much better from the multi-band modulation amplitudes than the transit depths, considering the quality of our datasets. 

\begin{table*}
\centering
\caption{Priors and derived parameters of the transit analyses. $\mathcal{U}$(A, B) indicates the uniform prior from A to B and $\mathcal{N}$(X, Y$^2$) indicates the normal prior with the mean value X and standard deviation Y. \label{tab:trfit-results-eccentric}}
\begin{tabular}{lcc}
Parameter & Prior & Derived Value \\
\hline
\multicolumn{3}{l}{\it Primary transit parameters$^a$} \\
$T_0$ [BJD] & $\mathcal{N}$(2458515.64206, 0.00010$^2$) & $2458515.642139 \pm 0.000095$ \\
$P$ [days] & $\mathcal{N}$(3.48456498, 0.00000060$^2$)& $3.4845650 \pm 0.0000005$ \\
$\sqrt{e} \cos \omega$ & $\mathcal{U}$(-1, 1) & $-0.82_{-0.04}^{+0.11}$  \\
$\sqrt{e} \sin \omega$ & $\mathcal{U}$(0, 1) & $0.12_{-0.09}^{+0.18}$  \\
$a/R_\star$ & $\mathcal{U}$(0, 100) &  $20.8 \pm 0.8$\\
$b$ & $\mathcal{N}$(0.628, 0.04$^2$) &   $0.65 \pm 0.02$\\
$R_p/R_\star$~($g$)& $\mathcal{U}$(0, 1) & $0.090_{-0.010}^{+0.009}$\\
$R_p/R_\star$~($r$)& $\mathcal{U}$(0, 1) & $0.104 \pm 0.004$\\
$R_p/R_\star$~($i$)& $\mathcal{U}$(0, 1) & $0.108 \pm 0.002$\\
$R_p/R_\star$~($z$)& $\mathcal{U}$(0, 1) & $0.108 \pm 0.002$\\
$R_p/R_\star$~(TESS)& $\mathcal{U}$(0, 1) & $0.101 \pm 0.005$\\
\multicolumn{3}{l}{\it Derived parameters and additional constraint} \\
$e$ & - & $0.69_{-0.10}^{+0.05}$ \\
$\omega$ [deg] & - & $172_{-14}^{+6}$ \\
$\rho_\star$ [g/cm$^3$] & $\mathcal{N}$(14.7, 1.5$^2$) &   $14.1 \pm 1.5$ \\
\hline
\multicolumn{3}{l}{\it Limb darkening parameters$^b$} \\
$q_1$~($g$) & $\mathcal{N}$(0.729, 0.011$^2$) & $0.721 \pm 0.012$ \\
$q_2$~($g$) &  $\mathcal{N}$(0.340, 0.002$^2$) & $0.343 \pm 0.003$ \\
$q_1$~($r$) & $\mathcal{N}$(0.633, 0.012$^2$) & $0.631 \pm 0.013$ \\
$q_2$~($r$) &  $\mathcal{N}$(0.366, 0.003$^2$) & $0.367 \pm 0.003$ \\
$q_1$~($i$) &  $\mathcal{N}$(0.430, 0.008$^2$) & $0.421 \pm 0.008$ \\
$q_2$~($i$) &  $\mathcal{N}$(0.294, 0.003$^2$) & $0.298 \pm 0.003$ \\
$q_1$~($z$) &  $\mathcal{N}$(0.337,0.007$^2$) & $0.322 \pm 0.007$ \\
$q_2$~($z$) &  $\mathcal{N}$(0.253,0.003$^2$) & $0.262 \pm 0.003$ \\
$q_1$~(TESS) &  $\mathcal{N}$(0.299,0.004$^2$) & $0.299 \pm 0.004$ \\
$q_2$~(TESS) &  $\mathcal{N}$(0.295,0.002$^2$) & $0.295 \pm 0.002$ \\
\hline
\multicolumn{3}{l}{\it Gaussian process hyperparameters} \\
$\log \sigma$~($g$) & $\mathcal{U}$(-10, 5) & $-3.8 \pm 0.3$ \\
$\log \rho$~($g$) & $\mathcal{U}$(-10, 10) & $-2.9 \pm 0.3$ \\
$\log \sigma$~($r$) &$\mathcal{U}$(-10, 5) & $-4.2 \pm 0.3$  \\
$\log \rho$~($r$) & $\mathcal{U}$(-10, 10) & $-2.8 \pm 0.3$  \\
$\log \sigma$~($i$) & $\mathcal{U}$(-10, 5) & $-5.5 \pm 0.4$  \\
$\log \rho$~($i$) &  $\mathcal{U}$(-10, 10) & $-2.4 \pm 0.4$  \\
$\log \sigma$~($z$) &  $\mathcal{U}$(-10, 5) & $-6.6 \pm 0.1$  \\
$\log \rho$~($z$) &  $\mathcal{U}$(-10, 10) & $-4.2 \pm 0.3$  \\
$\log \sigma$~(TESS) &  $\mathcal{U}$(-10, 5) & $-6.3 \pm 0.2$  \\
$\log \rho$~(TESS) &  $\mathcal{U}$(-10, 10) & $-3.4 \pm 0.5$  \\
\hline
\multicolumn{3}{l}{$^a$ For parameters with normal prior distributions, the mean and variance are adopted from \citet{Stefansson2020}}\\
\multicolumn{3}{l}{$^b$ from [$u_1, u_2$] values calculated by \ldtk .}\\
\end{tabular}
\end{table*}

\section{Discussion} \label{sec:discussion}

\subsection{Characteristics of spots on K2-25}\label{sec:dis-spots}

From the amplitudes of stellar brightness variations (Sections~\ref{sec:an-amplitude} and \ref{sec:an-spottemp}), we derived the spot temperature $T_{\rm spot}$ to be $\Delta T<$\upperT \,K lower than the assumed photosphere temperature of $T_{\rm phot}=3200$\,K, and the spot covering fraction $f_{\rm spot}$ to be $<$\upperf \% with $2\sigma$ confidence level (Figure~\ref{fig:corner-simple}). 
While the upper limit of spot covering fraction $f_{\rm spot}<\upperf\%$ is not a strong constraint, the upper limit of spot temperature, $\Delta T_{\rm spot} <\upperT$K, provides sufficient information to scrutinize spot properties and assess the extent of the TLSE, as detailed in Section~\ref{sec:dis-atmosphere}.

Furthermore, we explored the potential spot distribution through spot mapping (Section~\ref{sec:an-distribution}). Despite limiting the number of spots up to five due to parameter degeneracies, we determined that cases involving three or more spots exhibited a general trend in spot distribution. A further increase in the number of spots could elevate the value of $f_{\rm spot}$ due to certain spots not contributing to brightness variations (i.e., always visible spots). Nonetheless, $\Delta T$ remained relatively stable as the number of spots increased, indicating minimal variation in $\Delta T$ prediction with different numbers of spots.

Comparing the results from the amplitude (Figure~\ref{fig:corner-simple}) and spot mapping (Figure~\ref{fig:corner_spotnumber}), the spot mapping results showed more strict constraints on both the spot temperature and spot covering fraction in each spot number case. This is because the spot mapping takes into account not only the amplitude of the brightness variation but also the shape of the light curves. Possible sizes and positions of the spots were determined in order to represent the smooth brightness variations of K2-25 monitoring light curves, which resulted in a stronger constraint on the spot temperatures. We note that spot mapping entails various assumptions (e.g., assuming a perfectly circular spot, a small number of spots, etc.). The results may vary if a more flexible spot configuration is allowed \citep[e.g.,][]{Luger2021_mapping2}. Therefore, the more conservative results obtained from the amplitudes are used in the discussion part of this paper.

In addition, from the wavelength dependence of the transit depths (Section~\ref{sec:an-tlsetransit}), we also derived the $2 \sigma$ upper limit of both spot temperature difference ($<$1184\,K) and covering fraction ($<$0.43) (Figure~\ref{fig:corner_transit}). Despite weak constraints due to transit depth uncertainties, we can rule out the extreme cases with large spot temperature differences and large spot covering fractions. The results are consistent with the results derived by stellar brightness monitoring. Consequently, the resulting spot characteristics were in agreement with the overall features from the modulation amplitudes (Section~\ref{sec:an-spottemp}), spot mapping (Section~\ref{sec:an-distribution}) and from transit depths (Section~\ref{sec:an-tlsetransit}). 

\subsection{Effectiveness of simultaneous multi-band observations} \label{sec:effective}
In our study, simultaneous multi-band monitoring observations allowed us to achieve exceptional precision to constrain the spot characteristics, especially spot temperature difference $\Delta T$. This is mainly because, thanks to the multi-band light curves, degeneracies between spot latitude, spot size and spot temperature are broken to some extent. When the spot is located near the poles, the wavelength dependence of the dimming due to the spot is smaller because of limb-darkening. In this study, the wavelength dependence of the brightness variability amplitudes resulted in taking a low likelihood value of the case with spots near the pole, thereby better defining the spot size and temperature.

We confirmed that our method using multi-band light curves from TESS and ground-based telescopes resulted in better constraint especially on spot temperature estimation, comparing the spot mapping result using five light curves and using only the TESS light curve (details are in Appendix~\ref{sec:ap-multi}). The result of spot mapping effectively depends on the TESS light curve with higher time cadence and photometric precision, but multi-band data from ground-based telescopes are useful in putting further constraints on the solutions. 

There are a few studies of such multi-band (but not simultaneous) monitoring observations conducted for M-dwarfs. \citet{Morris2018} analysed the monitoring light curves of an M8 star TRAPPIST-1, by Kepler and Spitzer, and constrained the spot temperature to be extremely hot ($T_{\rm spot}\gtrsim 5300\pm200$\,K). \citet{Miyakawa2021} analysed the monitoring light curves of four M dwarfs in the Hyades cluster in the Kepler-band and $J, H, K$-band, and derived their spot temperatures within $1\sigma$ uncertainties of $\lesssim 700$\,K.

Our observations constrained the $T_{\rm spot}$ values more precisely than these studies. This is because our observations obtained multi-band light curves in optical, which are more sensitive to spot temperature than in infrared. Also, simultaneity was important, as the spot characteristics may change in a timescale of the spot emergence and decay (Appendix~\ref{sec:ap-depvar}).

Our observations showed that simultaneous multi-band monitoring using a combination of ground-based and space-based telescopes will be effective in determining the ``current" spot characteristics of the host star, for example, in future planetary atmosphere research.

\subsection{Indication to the spot-photosphere temperature relations} \label{sec:spot-photosphere}

\begin{figure*}
  \centering
      \includegraphics[width=0.9\textwidth]{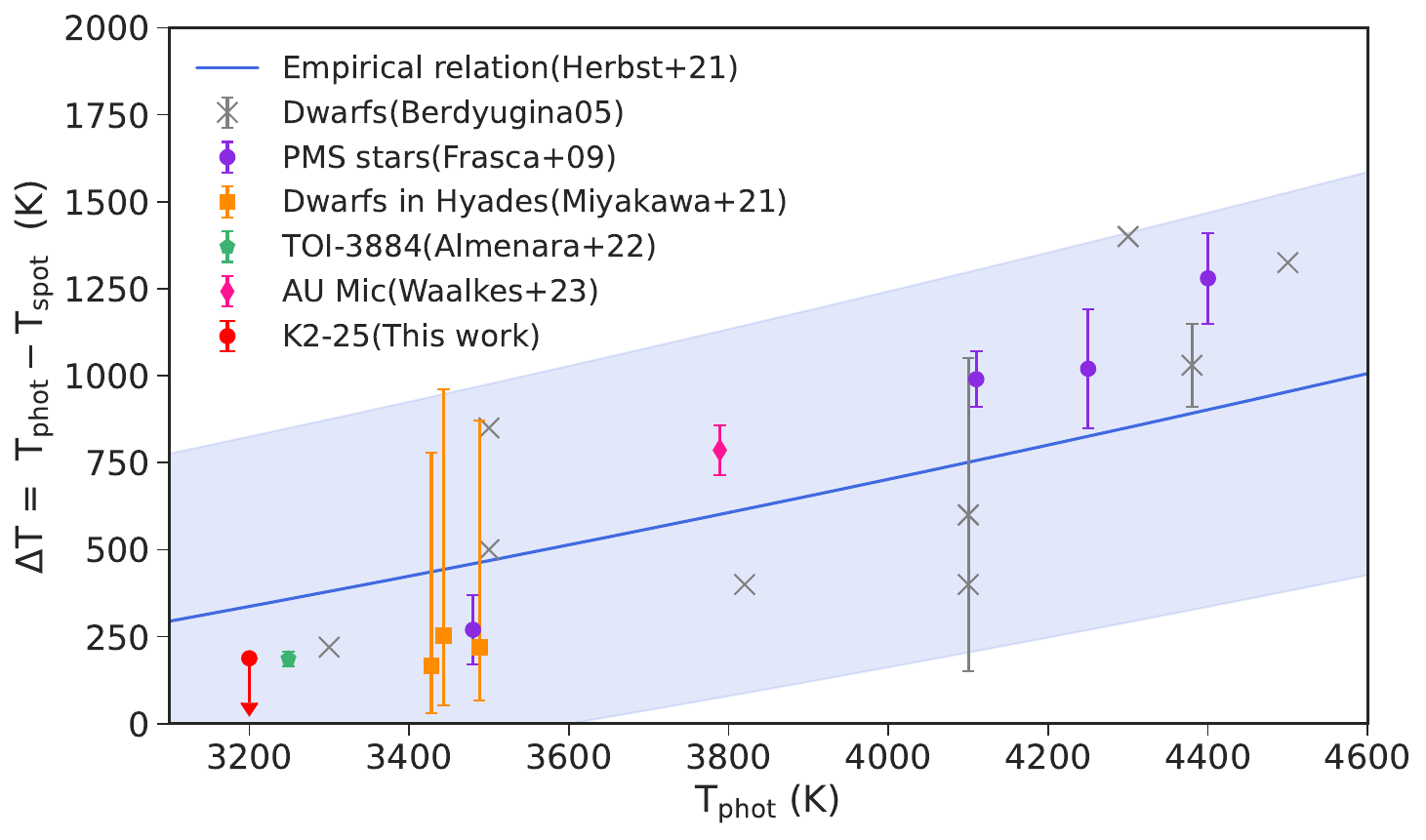}
  \caption{Relations between $T_{\rm phot}$ and temperature difference $\Delta T_{\rm spot} = T_{\rm phot} - T_{\rm spot}$ for the dwarfs between 3000\,K to 5000\,K. The empirical relation and its uncertainties from Equation~4 in \citet{Herbst2021} (derived from dwarfs and giants in \citet{Berdyugina2005}) are indicated in the blue line and filled regions. Note that data points without error bars are those for which the errors are not estimated. For the stars with $T_{\rm phot}\lesssim 3500$\,K, $\Delta T$ tend to show smaller values compared to the empirical relation.} 
  \label{fig:Tspotrelation}
 \end{figure*}

Figure~\ref{fig:Tspotrelation} shows the relation of $\Delta T_{\rm spot} = T_{\rm phot} - T_{\rm spot}$ and $T_{\rm phot}$ values from this work in comparison with the ones from recent literature for cool dwarfs derived by various methods \citep{Berdyugina2005, Frasca2009, Miyakawa2021, Almenara2022, Waalkes2023}.

There were several studies to derive the empirical relations of $\Delta T$ and $T_{\rm phot}$ with polynomial functions \citep{Berdyugina2005, Maehara2017, Herbst2021}. \citet{Berdyugina2005} reports that $\Delta T$ increases as the value of $T_{\rm phot}$ increases for both the dwarfs and giants. \citet{Herbst2021} updates the empirical relations by incorporating the latest samples, mainly for stars with effective temperatures of $T_{\rm eff}>4500$\,K.
After \citet{Berdyugina2005}, a few observations slightly increased the number of samples with $T_{\rm eff} < 4000$\,K \citep{Frasca2009, Miyakawa2021, Almenara2022, Waalkes2023}. The results of those studies and our result suggest that the overall trend almost corresponds to the empirical relationship, but a trend towards smaller $\Delta T$ with lower photosphere temperature, except for a young pre-main sequence M dwarf AU Mic \citep{Waalkes2023}.

Whereas the number of samples is still limited, these results suggest that the relations might no longer be suitable for M-dwarfs, especially with $T_{\rm eff} \lesssim 3500$\,K. It is important to note that in the prediction of the TLSE for each planetary system, $T_{\rm spot}$ values are often assumed based on the empirical formula mentioned above \citep[e.g.,][]{Rackham2018}. In such cases, there is a possibility of overestimating $\Delta T$ values (i.e., underestimating $T_{\rm spot}$ values). It is possible to update the relationship for M-dwarfs by adding samples with high-precision measurements of $\Delta T$ as future issues. In addition, the accumulation of the measurements on the $T_{\rm spot}$ for low-temperature stars would be useful in theoretical studies of the starspot properties. In \citet{Panja2020}, the temperature dependence of the $\rm{H^{-}}$ opacity explains the trend of $\Delta T_{\rm spot}$ and $T_{\rm phot}$, and this reproduces the trend derived by \citet{Berdyugina2005}. However, the calculation has not yet been done for stars with temperatures lower than 3700 K (i.e., M0-type stars). It is important to increase observational data for the evaluation of these theoretical explanations.

\subsection{Indication to the K2-25b atmosphere retrievals}\label{sec:dis-atmosphere}

\begin{figure*}
    \centering
    \includegraphics[width=1.0\textwidth]{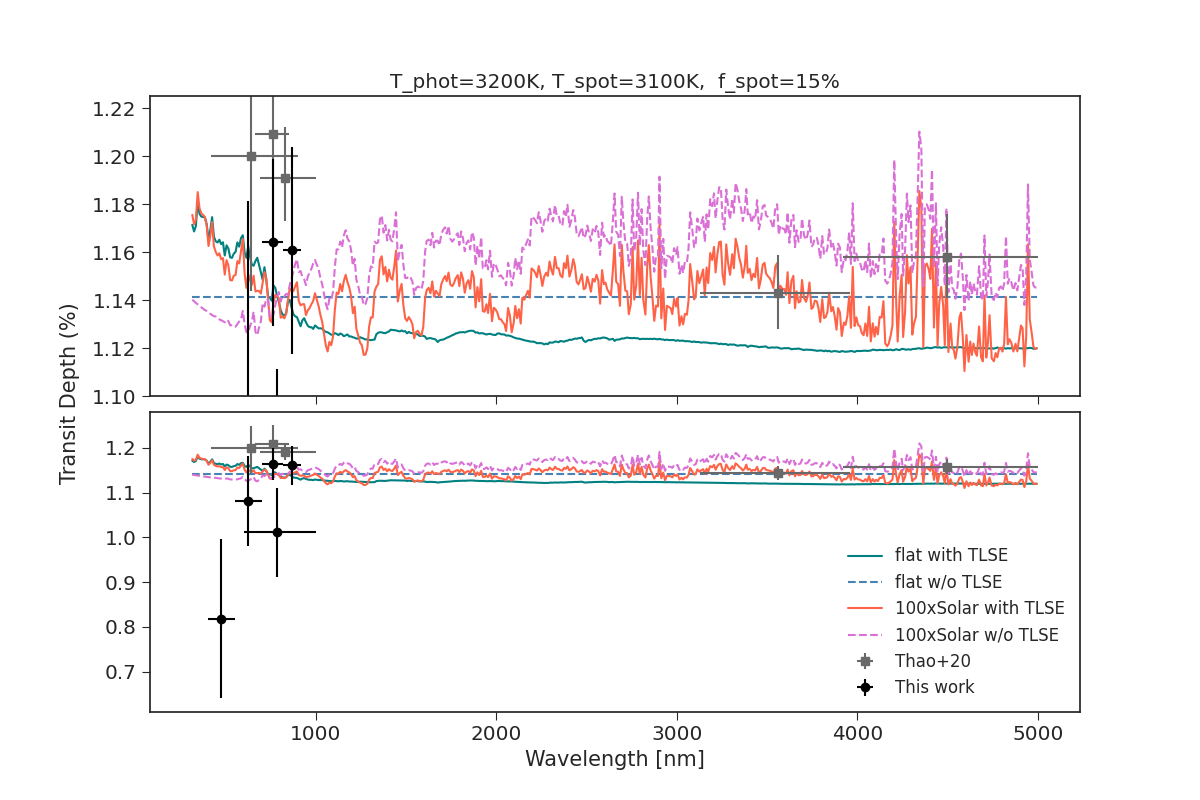}
    \caption{Generated transmission spectrum of K2-25b overplotted with the observed transit depths in multi-band. The black dots with errorbars indicate the transit depths by this work in $g$, $r$, $i$, TESS and $z$-band from left to right, and grey dots with errorbars indicate the values from \citet{Thao2020}. The observed transit depths are compared with the planetary atmosphere models. Green/blue lines are the flat atmosphere model and orange/pink lines are the forward model generated by \platon, with $100\times$ solar metalicity atmosphere, with (solid) and without (dashed) the effect of spots (TLSE). The y-axis offset of each model was optimized for the data obtained in this work.} The top figure is the close-up version of the bottom figure.
    \label{fig:platon}
\end{figure*}

In Figure~\ref{fig:platon}, we show the derived transit depths of K2-25\,b in five bands ($g$, $r$, $i$, $z$, and TESS-band) along the wavelength. We overplotted them with the previous transit depths derived by \citet{Thao2020}, which used the data taken with Kepler, Sinistro $i$-band, MEarth, Spitzer channel 1, and Spitzer channel 2. We found that the transit depths from our observations have smaller overall values than the ones from \citet{Thao2020}, despite the wavelength ranges being similar. However, given the large errors especially in the $g$, $r$, and TESS bands, all transit depths are consistent within a $2\sigma$ range. We did not ascertain the trend of deeper transit in the visible compared to the infrared, which was reported in \citet{Thao2020}. This result might be due to some systematics from different observations, different analytical methods, and different stellar activity levels.

We also overlaid the calculated transmission spectrum models assuming TLSE and planetary atmosphere compositions. The atmosphere model with 100$\times$ solar metallicity is calculated using the python package \platon \citep{platon} assuming the chemical equilibrium, from $R_\star = 0.2932R_\odot$ \citep{Thao2020}, $R_p/R_s=0.108$ (this work), $M_p=24.5 M_\oplus$ and the equilibrium temperature $T_{\rm eq}=494$\,K \citep{Stefansson2020}. The TLSE was calculated assuming the spots with $T_{\rm spot} = 3100$\,K and $f_{\rm spot} = 15$ \,\%, which are reasonable values from our monitoring observations. 

While changing the y-axis offset as a parameter, we fit the models to data and calculated the chi-squared and reduced chi-squared values for four models: a flat atmosphere spectrum with and without TLSE, and a 100$\times$-Solar-metalicity atmosphere spectrum with and without TLSE. The values are listed in Table~\ref{tab:chi2_tlse}. We do not see a significant difference in chi-squared values for all models, in both cases using the transit depths only from this work and from this work combined with \citet{Thao2020}.

While \citet{Thao2020} rejected the clear $100\times$ solar metalicity model, our analysis shows that this possibility cannot be dismissed. This is not due to differences in stellar surface models, but mainly because the atmospheric model features were reduced due to the larger planetary mass estimates, with reference to the results of \citet{Stefansson2020}. For the combined dataset from our work and \citet{Thao2020}, the chi-squared value is slightly smaller when the spot effects are taken into account, for both the flat and 100$\times$-Solar-metalicity atmosphere cases, but this is also not a statistically significant difference. 

Also, in \citet{Thao2020}, they pointed out that spot covering fractions of 22\% to 36\% are too large and unlikely given that the stellar variability is 0.5\% to 2\%, but our analysis does not rule out such a large spot covering fraction, considering the existence of spots that do not contribute to the stellar brightness variability.

It should be noted that the spectrum from \citet{Thao2020} is stacked, i.e. consisting of individual datasets taken between 2015 and 2017, while our spectrum was taken in 2021. Given that our observations show that the spot distribution in K2-25 varies on scales of hundreds of days (see Appendix~\ref{sec:an-longterm}), it would be not good practice to use all datasets to compare with models. We therefore stop here for a detailed comparison using both results. To make a real comparison for the atmosphere constraint of K2-25b, we need to simultaneously obtain spectra over a wide range of wavelengths with high-precision spectroscopy, such as by JWST. 

As in Figure~\ref{fig:varepsilon}, the TLSE is not very significant in the JWST/NIRSPEC wavelength range (600 - 5000\,nm) compared to the optical. However, for example, when comparing the contamination factor at wavelength $\sim$ 1500\,nm and 4000\,nm, there are some cases where the $\epsilon_\lambda$ differs by $\sim 1$ \%. This means that even if the ``true" transit depth is the same (i.e., flat transmission spectra), the observed depth differs by 100\,ppm relative to each other. 

Given the photometric precision of JWST's observations, this effect cannot be ignored. We have confirmed from the simulation of JWST spectra using software \texttt{PandExo} \citep{PandExo}, that when K2-25 is observed with PRISM at JWST/NIRSPEC, a precision of $ <100$ ppm can be achieved in a single transit, although this depends on the wavelength binning. This indicates that stellar surface models need to be examined when inferring atmospheric models for K2-25\,b from transmission spectroscopy by JWST. In that case, our knowledge of the spot temperature and the covering fraction of K2-25 from the monitoring observations is a good prior to investigating the stellar surface models. Of course, ideally, the monitoring observations should be conducted at the same time as the transmission spectroscopy by JWST.

\begin{table}
\centering
\caption{Calculated chi-squared values and reduced chi-squared values for each atmosphere model with and without spot effect, assuming $T_{\rm spot} = 3100$\,K and $f_{\rm spot} = 15$ \,\%. The values are calculated for (1) only using MuSCATs and TESS data from this work and (2) combined data from this work and from \citet{Thao2020} (T20).}\label{tab:chi2_tlse}
\begin{tabular}{lcccc}\hline
Model & \multicolumn{4}{c}{Data} \\
 & \multicolumn{2}{c}{(1)This Work} & \multicolumn{2}{c}{(2)This Work + T20} \\\hline
 & \multicolumn{1}{l}{$\chi^2$} & \multicolumn{1}{l}{reduced $\chi^2$} & \multicolumn{1}{l}{$\chi^2$} & \multicolumn{1}{l}{reduced $\chi^2$} \\\hline
Flat with spot & 6.82 & 1.70 & 10.6 & 1.17 \\
Flat without spot & 5.98 & 1.49 & 11.5 & 1.28 \\
$100 \times$Solar with spot & 6.50 & 1.62 & 13.1 & 1.46 \\
$100 \times$Solar without spot & 5.66 & 1.42 & 17.2 & 1.91\\\hline
\end{tabular}
\end{table}

\subsection{Other stellar activity features on K2-25: indication for the future atmosphere research}

We also investigated some other stellar activity signals on K2-25 from the analyses of multi-band light curves (see Appendix~\ref{sec:an-longterm}, \ref{sec:ap-depvar} to \ref{sec:an-flare} for details).
We found there is a long-term variability of the modulation amplitude in a timescale of a few $\times 100$ days. This might be explained by the timescale of the change in the spot distribution. This means it is important to obtain the concurrent status of stellar activity when the transmission spectroscopy is conducted. 

From the analysis of flare frequency, we confirmed three flares in K2-25 with the energy at $\sim 10^{33}$ erg, in a $21.76$ days-long TESS light curve. This is a typical flare frequency for a young mid-M star \citep{Gunther2020, Murray2022}, although the number of flare detections was limited because of the detection limit. Smaller flares should happen more frequently, which could be the dominant source of baseline fluctuation in transit light curves (especially in the blue-colour band). When performing high-precision transit observations using space telescopes, at least multiple observations should be made, since flares may occur during the observation and cause the baseline modulation in the transit light curves.

\section{Conclusion} \label{sec:coclusion}

In this study, we examined the stellar brightness variability of the young mid-M-dwarf K2-25 and analyzed the transits of its sub-Neptune companion K2-25\,b using data from TESS and ground-based telescopes. Our aim was to investigate the characteristics of starspots within the system and their impact on transit observations.

The significance of our study stems from the utilization of (i,ii) ``multi-band" and (iii) ``simultaneous" observations, which enabled precise constraints on spot characteristics.

\begin{itemize}
\item[(i)]{Multi-band monitoring of stellar brightness provided robust constraints on spot temperature. Using monitoring light curves in four different bands (utilizing five different instruments/filters), we revealed the wavelength dependence of the amplitudes of stellar brightness modulation. From there, we found the spot temperature difference from the photosphere temperature of K2-25 to be $<$\upperT\,K and the spot covering fraction to be $<$\upperf \% with $2\sigma$ confidence level. These results were consistent with the one derived from spot mapping, assuming two, three, four or five spot cases (Section~\ref{sec:dis-spots}).
The derived spot temperature difference was smaller than the extrapolation of the empirical relation with the stellar photosphere temperature. This suggests that such an empirical relation is no longer suitable for M-dwarfs (Section~\ref{sec:spot-photosphere}).}

\item[(ii)] {Multi-band transit photometry was also obtained to constrain stellar surface models through the calculation of the TLSE. Although the derived transit depths by MuSCAT series and TESS had uncertainties too large to precisely constrain the spot temperature and covering fraction, we confirmed that the results were consistent with the spot characteristics constrained by the monitoring observations (Section~\ref{sec:an-tlsetransit}). We calculated that the expected TLSE would distort the transmission spectrum of the system with the amplitude of $\sim 100$ ppm in the JWST/NIRSPEC wavelength range, which is detectable with JWST (Section~\ref{sec:dis-atmosphere})}. 

\item[(iii)] {This study was the first trial of taking stellar monitoring and transit observations simultaneously (Section~\ref{sec:effective}). From the long-baseline light curves, we found that the spot distributions should change on timescales of hundreds of days. For more precise transit observations, it is important to know the spot distribution at each observing epoch. }
\end{itemize}

There is room for further improvements to our methods, such as improving observation precision and refining stellar surface models for more precise derivation of stellar parameters from light curves. For example, faculae were not considered in our analysis because of the difficulty in quantifying their contributions to the light curves: faculae are known to become brighter towards the star limb \citep{Norris2023}. Although faculae are expected to contribute less to the brightness variability than spots because faculae are more widely distributed on the stellar surface \citep{Johnson2021}, how to handle the contribution from faculae to the light curve is a future issue.

Nevertheless, our investigation of the spot characteristics of an M-dwarf, using the best  current instruments and methods, is an important step forward. Employing this method, along with further improvements as mentioned above, may be crucial for more accurate observations of planetary atmospheres with space telescopes, and for further studies of the stellar activity of M-dwarfs.

\section{Data Availability} 
The TESS data can be accessed through the MAST (Mikulski Archive for Space Telescopes) portal at \url{https://mast.stsci.edu/portal/Mashup/Clients/Mast/Portal.html}. The ZTF data can be accessed through 
\url{https://irsa.ipac.caltech.edu/Missions/ztf.html}.
The data taken by Sinistro, MuSCAT2 and 3 are available upon request to the authors. The SVO Filter Profile Service and theoretical stellar spectra can be obtained from the SVO Theoretical Model Services \url{http://svo2.cab.inta-csic.es/theory/main/}.

This work has utilized these public software: LCOGT \texttt{BANZAI} pipeline \citep{McCully2018}, \texttt{PyPeriod} \citep{pya}, \corner \citep{corner}, \celerite \citep{celerite1,celerite2}, \batman \citep{BATMAN}, \ldtk \citep{Parviainen2015_ldtk,Husser2013}, \emcee \citep{emcee},  \platon \citep{platon}, \texttt{PandExo} \citep{PandExo}, \texttt{Llamaradas-Estelares} \citep{Mendoza2022}, and \texttt{Astropy} \citep{astropy:2013,astropy:2018}. 

\section{Acknowledgements}

We acknowledge Dr.~Teruyuki Hirano and Dr.~Kohei Miyakawa for their helpful insights and advice during the discussion process. 

We acknowledge the use of public TESS data from pipelines at the TESS Science Office and at the TESS Science Processing Operations Center. Funding for the TESS mission is provided by NASA's Science Mission Directorate.  
This research has made use of the Exoplanet Follow-up Observation Program website, which is operated by the California Institute of Technology, under contract with the National Aeronautics and Space Administration under the Exoplanet Exploration Program. Resources supporting this work were provided by the NASA High-End Computing (HEC) Program through the NASA Advanced Supercomputing (NAS) Division at Ames Research Center for the production of the SPOC data products.

This work is partly based on observations made with the MuSCAT2 instrument, developed by ABC, at Telescopio Carlos S\'{a}nchez operated on the island of Tenerife by the IAC in the Spanish Observatorio del Teide.

This work makes use of observations from the Las Cumbres Observatory global telescope network. Some data in the paper are based on observations made with the MuSCAT3 instrument, developed by Astrobiology Center and under financial support by JSPS KAKENHI (JP18H05439) and JST PRESTO (JPMJPR1775), at Faulkes Telescope North on Maui, HI, operated by the Las Cumbres Observatory.

Based on observations obtained with the Samuel Oschin Telescope 48-inch and the 60-inch Telescope at the Palomar Observatory as part of the Zwicky Transient Facility project. ZTF is supported by the National Science Foundation under Grant No. AST-2034437 and a collaboration including Caltech, IPAC, the Weizmann Institute for Science, the Oskar Klein Center at Stockholm University, the University of Maryland, Deutsches Elektronen-Synchrotron and Humboldt University, the TANGO Consortium of Taiwan, the University of Wisconsin at Milwaukee, Trinity College Dublin, Lawrence Livermore National Laboratories, and IN2P3, France. Operations are conducted by COO, IPAC, and UW.

This work is supported by Grant-in-Aid for JSPS Fellows, Grant Number JP20J21872, 
JSPS KAKENHI Grant Numbers 
JP18H05439, 
18H05442, 
21K13955, 
21K20376, 
and JST CREST Grant Number JPMJCR1761. 

We acknowledge financial support from the Agencia Estatal de Investigaci\'on of the Ministerio de Ciencia e Innovaci\'on MCIN/AEI/10.13039/501100011033 and the ERDF ``A way of making Europe" through project PID2021-125627OB-C32, and from the Centre of Excellence ``Severo Ochoa" award to the Instituto de Astrofisica de Canarias.

\bibliographystyle{mnras}
\bibliography{bibtex}{}

\clearpage
\appendix
\renewcommand\thefigure{\thesection.\arabic{figure}}
\renewcommand\thetable{\thesection.\arabic{table}}

\section{Additional Analyses}\label{sec:addition}

\subsection{Analysis on the blended sources in the TESS photometry aperture}\label{sec:ap-aperture}
As K2-25 is located in a relatively crowded field, its photometric aperture of TESS could be contaminated by other sources. Since the blended sources can modify the observed brightness variability, we examined the effect of their contamination.

\begin{figure}
  \centering
      \includegraphics[width=0.45\textwidth]{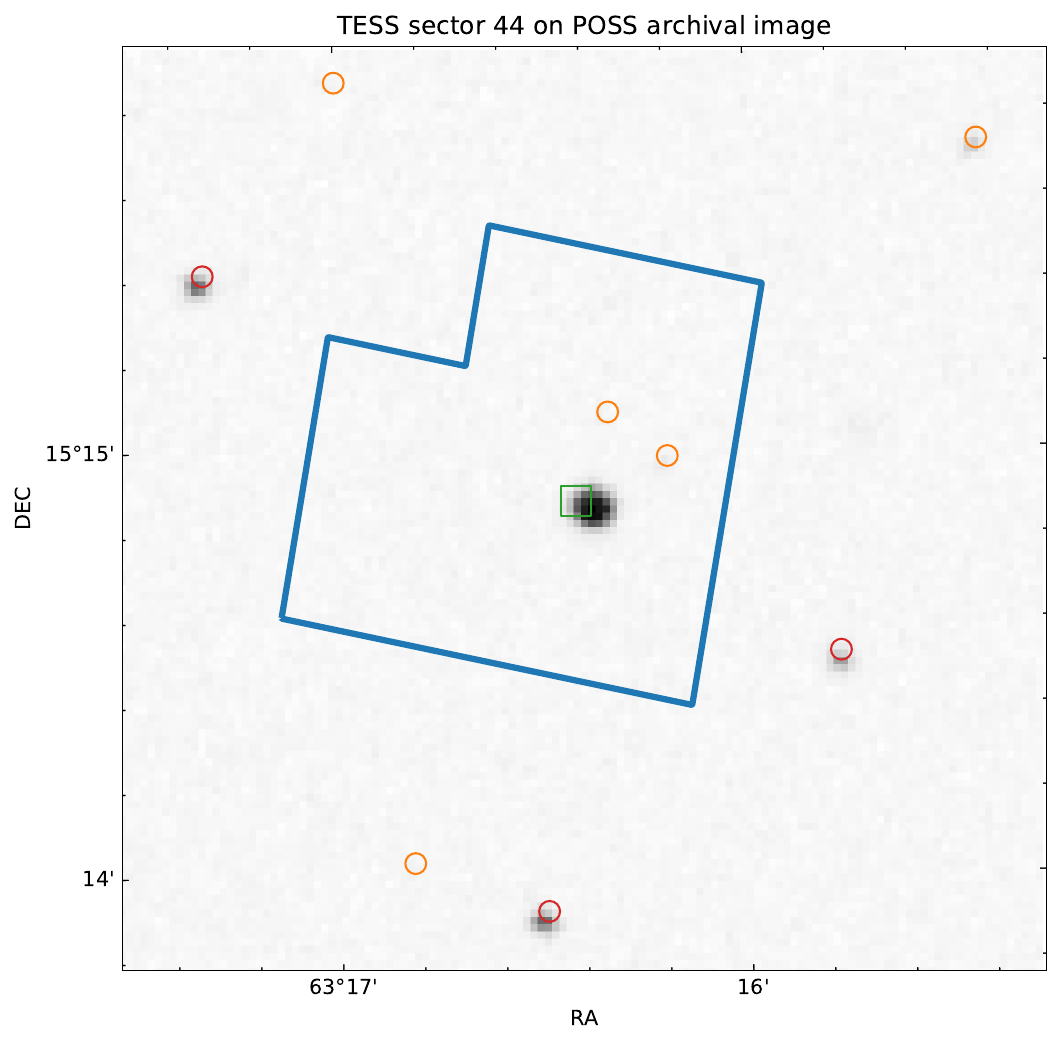}
  \caption{Archival imaging from POSSII-F survey \citep{POSII1991} with the TESS photometric aperture (blue outline) from Sector 44 and Gaia sources (red circles) from Gaia DR3\citep{Gaia2016,GaiaDR3_2023}. The green square indicates the position of K2-25.}
  \label{fig:tessap}
\end{figure}

Figure~\ref{fig:tessap} shows the TESS photometric aperture of K2-25, overplotted with the archival field image. There are two faint Gaia sources (Gaia 3311804511206599168 and 3311804515501995648) in the aperture. Both of them are $\sim 6$\,mag fainter than K2-25 in Gaia band \citep[Gaia DR3;][]{Gaia2016,GaiaDR3_2023}, which means they contribute only $<0.5\%$ of the measured flux. They can only change $\lesssim 1\%$ of the measured amplitude of the brightness variation, which is within the uncertainty of the amplitude measured with TESS (\ref{tab:amplitudes}). Therefore, we concluded that the effect of the blended sources is negligible. 

\subsection{Another way to derive modulation amplitudes - Gaussian Process}\label{sec:ap-gp}
To check the robustness of the results of the method presented in Section~\ref{sec:an-amplitude} (we call ``sine-curve method" hereafter), we also obtained the brightness variation amplitudes of the five light curves in the following way, using the Gaussian process (GP). Assuming that the light curves are represented by a combination of periodic variations by the stellar rotation and white noise due to systematics, we used a combination of the modified quasi-periodic (MQP) kernel and the white-noise (WN) kernel in the Python package \celerite:
\begin{align}
    k(\tau)  &= k_{\rm MQP}(\tau) + k_{\rm WN}(\tau)\\
    k_{\rm MQP}(\tau) &= \frac{\theta_1}{2+\theta_2}e^{-\tau/\theta_3}\Big[\cos\big(\frac{2\pi \tau}{P}\big)+(1+\theta_2)\Big] \\
    k_{\rm WN}(\tau) &= \theta_4\delta(p, p') .
\end{align}
Here, $\tau = |\bm{p}-\bm{p'}|$ is the difference of two data points at the stellar rotation phase $\bm{p}$, and $\theta_1, \theta_2, \theta_3, \theta_4, P$ are the hyperparameters. $\theta_1$ and $\theta_2$ tune the strength of correlation between each observation point; $\theta_3$ tunes the timescale of exponential decay; $\theta_4$ accounts for the white noise level; $P$ corresponds to period of the modulation.

We assumed the light curve is different only in the amplitude for each band, and the timescales of the decay are largely consistent in all bands. Therefore, we set $\theta_2$ and $\theta_3$ to be common parameters for all light curves, whereas $\theta_1$ and $\theta_4$ are different parameters for each light curve. Also, as the light curves are already phase-folded, we fixed the value $P = 1$. Therefore, we optimize the five light curves simultaneously, with 12 hyperparameters in total.  
We put log-uniform prior distributions for all the hyperparameters, which only prevent the hyperparameters from becoming too large and unphysical. We used MCMC package \emcee for the parameter estimation. Parameters, priors, and derived values for each parameter are summarized in Table~\ref{tab:rotation-gp}.

\begin{figure*}
  \centering
      \includegraphics[width=1.0\textwidth]{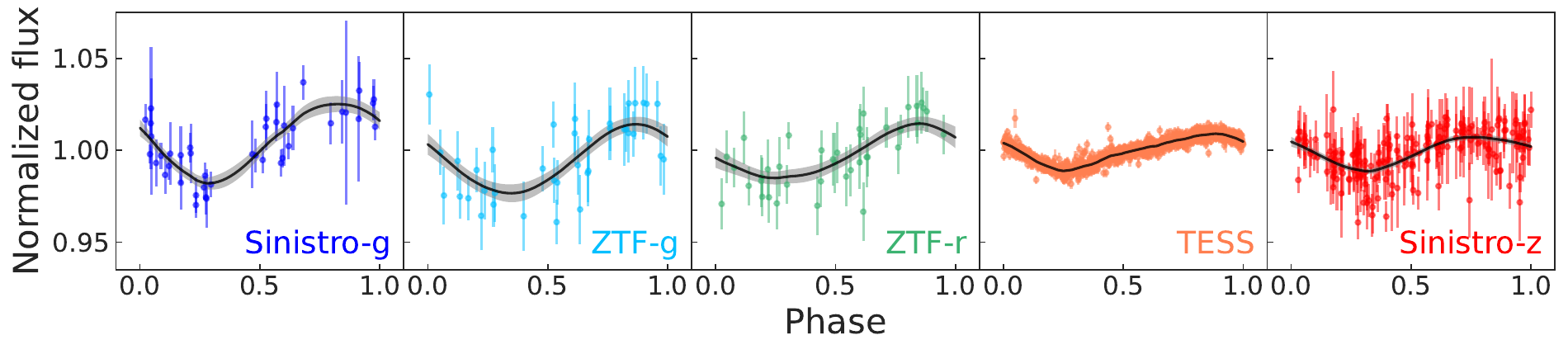}
  \caption{The phase-folded light curves and fitted model curves by the GP method. The solid line shows the best-fit curves and the shaded region shows the 1 $\sigma$ uncertainties of the models. The white noise are taken into account by adding white noise kernels.}
  \label{fig:folded-gp}
\end{figure*}

Figure~\ref{fig:folded-gp} shows the resulted GP models plotted with the observed light curves. The amplitudes and their errors are derived by calculating the difference between the maximum and minimum values of the resulting light curve model with errors.

The derived amplitudes from the sine-curve method and the GP method are compared in Table~\ref{tab:amplitudes}. The resulting posterior distribution of $\Delta T$ and $f_{\rm spot}$ is similar to the one by the sine-curve method. When adopting amplitude values from GP method, the median value and $2\sigma$-upper limit of $\Delta T$ and $f_{\rm spot}$ was 68\,K and 239\,K, and 16\% and 64\%, respectively. 

\begin{table}
\caption{Priors and derived parameters for joint GP fit for the modulation analysis. $\mathcal{U}$(X, Y) indicates the uniform prior which allows the range between X and Y.}\label{tab:rotation-gp}
\centering
\begin{tabular}{lcc}\hline
Hyperparameter                 & Prior                                     & Derived Value              \\\hline
$P$                            & 1 (fix)                                   & -                          \\
$\log \theta_1$ (Sinistro-$g$) & $\mathcal{U}$(-15, 3)                     & $-7.82_{-0.77}^{+0.95}$    \\
$\log\theta_1$ (ZTF-$g$)       & $\mathcal{U}$(-15, 3)                     & $-8.23_{-0.79}^{+0.90}$    \\
$\log\theta_1$ (ZTF-$r$)       & $\mathcal{U}$(-15, 3)                     & $-8.22_{-0.99}^{+1.08}$    \\
$\log\theta_1$ (TESS)          & $\mathcal{U}$(-15, 3)                     & $-9.82_{-0.62}^{+0.79}$    \\
$\log\theta_1$ (Sinistro-$z$)  & $\mathcal{U}$(-15, 3)                     & $-9.32_{-0.84}^{+0.94}$    \\
$\log\theta_2$                 & $\mathcal{U}$(-20, 5)                     & $-10.6_{-6.3}^{+6.2}$   \\
$\log\theta_3$                 & $\mathcal{U}$(-10, 5)                     & $-1.84_{-0.83}^{+0.68}$    \\
$\log\theta_4$ (Sinistro-$g$)  & $\mathcal{U}$(-10, 0)                     & $-4.66 \pm 0.14$           \\
$\log\theta_4$ (ZTF-$g$)       & \multicolumn{1}{l}{$\mathcal{U}$(-10, 0)} & $-7.0_{-2.0}^{+1.8}$       \\
$\log\theta_4$ (ZTF-$r$)       & \multicolumn{1}{l}{$\mathcal{U}$(-10, 0)} & $-5.00_{-0.74}^{+0.33}$       \\
$\log\theta_4$ (TESS)          & \multicolumn{1}{l}{$\mathcal{U}$(-10, 0)} & $-6.621_{-0.072}^{+0.067}$ \\
$\log\theta_4$ (Sinistro-$z$)  & \multicolumn{1}{l}{$\mathcal{U}$(-10, 0)} & $-4.584_{-0.054}^{+0.056}$\\\hline
\end{tabular}
\end{table}

\begin{table}
    \centering
     \caption{Derived modulation amplitude for the five light curves, by sine-method and GP method.}
    \begin{tabular}{c|ccc}\hline
        & \multicolumn{2}{c}{Amplitude (\%)} \\
        Band & Sine-curve method & GP method \\\hline
        Sinistro-$g$ & $4.16\pm0.48$ & $4.31\pm0.54$ \\
        ZTF-$g$ & $3.96\pm0.69$ & $3.75\pm0.63$\\
        ZTF-$r$ & $2.98\pm0.57$& $2.96\pm0.51$\\
        TESS & $1.954\pm0.022$& $2.029\pm0.055$\\
        Sinistro-$z$ & $2.00\pm0.21$ & $1.85\pm0.23$\\\hline
    \end{tabular}
    \label{tab:amplitudes}
\end{table}

\subsection{Analysis on the long-term change of the variability amplitude}\label{sec:an-longterm}

\citet{Kain2020} revealed that the modulation amplitude of K2-25 differs from year to year from the analyses of light curves observed with the MEarth Observatories \citep{Irwin2015} over a three-year period. To see the recent long-term variation of the modulation amplitude, we divided the ZTF light curves into 8 different segments, each of which has a 200-day period and derived modulation amplitudes by the GP method. This time, the hyperparameters were fixed to the derived value described in Table~\ref{tab:rotation-gp} for each ZTF band.

\begin{figure*}
  \centering
      \includegraphics[width=1.0\textwidth]{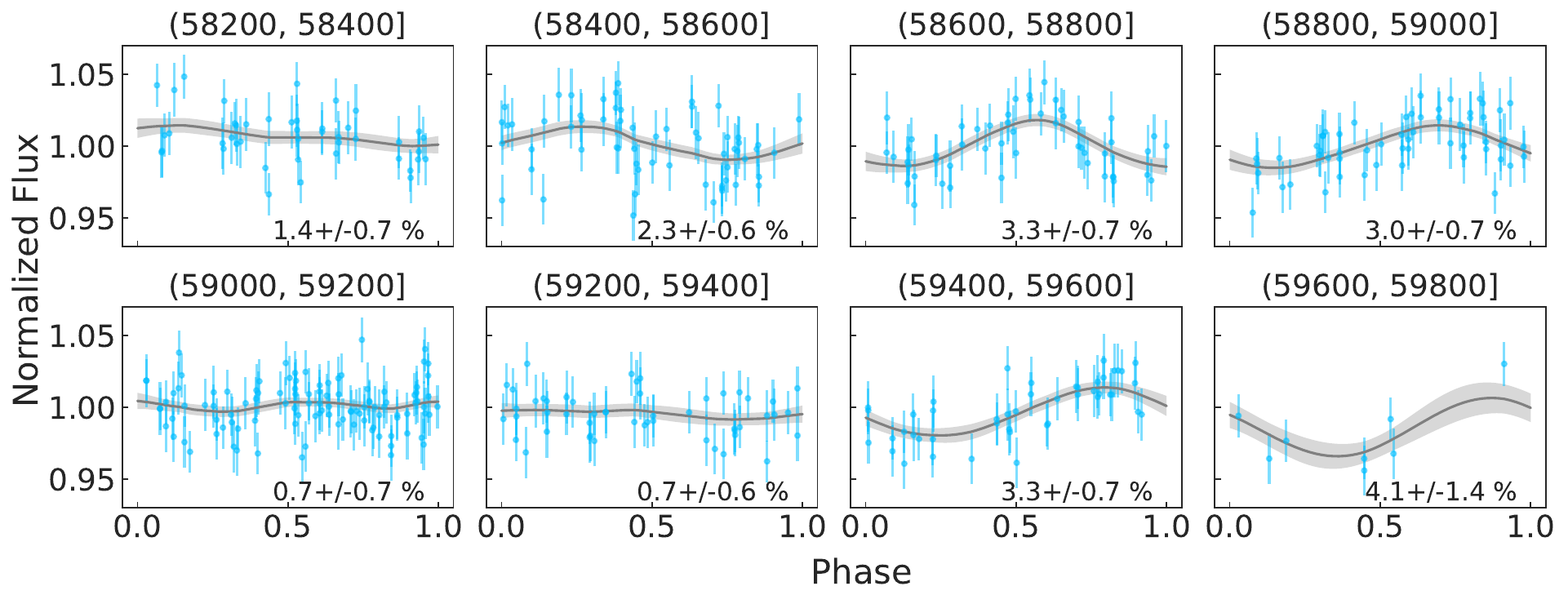}
      \includegraphics[width=1.0\textwidth]{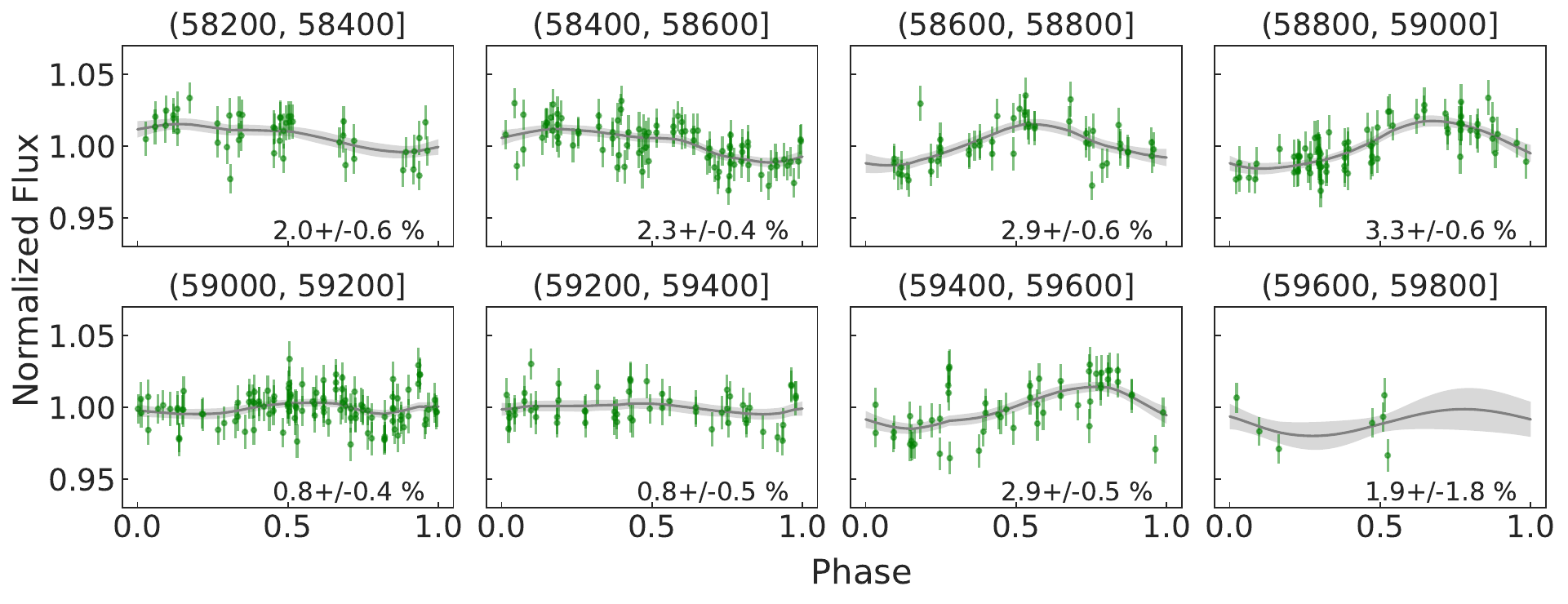}
  \caption{(top) Phase-folded light curves in $Z_g$-band by ZTF, divided into 8 bins every 200 days, with GP model. The title of each panel shows the range of MJDs of the data used for that plot. (bottom) Same figure but for $Z_r$-band data. The shape of the brightness modulation is different in each bin.}
  \label{fig:ztf_gr}
 \end{figure*}

The shape of the brightness modulation is changing in a timescale of $\sim$ 200\,days, as shown in Figure~\ref{fig:ztf_gr}. The amplitude also varied at each time, with a relatively large amplitude during the period we observed with other instruments (MJD 59400 - 59800), and almost no visible variations during the previous period (MJD 59000 - 59400). In addition, the shape of the variation is consistent between $Z_g$ and $Z_r$-band, with $Z_g$-band always showing a greater amplitude. 

This means the spot distribution has changed in a timescale of a few $\times 100$ days, and it is difficult to constrain spot characteristics by the combination of different observations at the different observing epochs.

\subsection{Possibility of bright spots}\label{ap:bright}
In this section, we deal with the case of bright spots, which was not considered in the derivation of spot temperature and covering fraction from the brightness modulation amplitudes (Section~\ref{sec:an-spottemp}). Bright spots can create similar modulations in light curves as described by Equation \ref{eq:fs}, although the bright spots are most visible when the star is brightest. We obtained a BT-Settl model with spot temperatures ranging from 3200\,K to 5000\,K in 100\,K increments, which was interpolated for use.

\begin{figure}
\centering
      \includegraphics[width=0.47\textwidth]{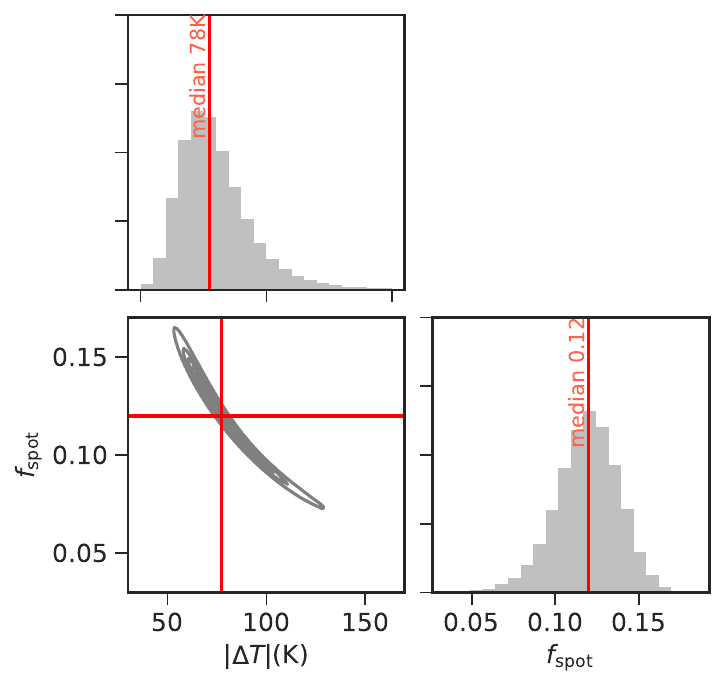}
  \caption{Similar to Figure~\ref{fig:corner-simple}, but for the case of bright spots. The red lines represent the median values of the posterior distributions.}
  \label{fig:corner-simple-faculae}
 \end{figure}

Figure~\ref{fig:corner-simple-faculae} shows the resulting posterior distribution of spot characteristics from the modulation amplitudes. The derived values are $|\Delta T_{\rm spot}| = 78 _{-11}^{+16}$\,K and $f_{\rm spot}=12\pm 2$\,\% with $1\sigma$ uncertainties. 

The TLSE caused by those bright spots results in smaller transit depths in shorter wavelength in the optical. We calculated the chi-squared values of the calculated transit depths to observed transit depths as done in Section~\ref{sec:dis-atmosphere}, adopting reasonable values for the bright spots case ($T_{\rm spot} = 3300$\,K and $f_{\rm spot} =15$\,\%). Table~\ref{tab:chi2_tlse_faculae} presents similar chi-squared values for all cases, which do not strongly suggest the existence of bright spots, considering the large errors in observed transit depths. 

Here, bright spots refer to regions that have a higher temperature than the photosphere, and their physical characteristics are not considered. These could be faculae, but as explained in the Conclusion, faculae exhibit complex behaviour in light curves and are challenging to implement in spot mapping. This remains a future task, including how such regions, which slightly warmer than their surroundings, can be physically explained.

 \begin{table}
\centering
\caption{Similar to Table~\ref{tab:chi2_tlse}, but assuming $T_{\rm spot} = 3300$\,K and $f_{\rm spot} = 15$ \,\%.} \label{tab:chi2_tlse_faculae}
\begin{tabular}{lcccc}\hline
Model & \multicolumn{4}{c}{Data} \\
 & \multicolumn{2}{c}{(1)This Work} & \multicolumn{2}{c}{(2)This Work + T20} \\\hline
 & \multicolumn{1}{l}{$\chi^2$} & \multicolumn{1}{l}{reduced $\chi^2$} & \multicolumn{1}{l}{$\chi^2$} & \multicolumn{1}{l}{reduced $\chi^2$} \\\hline
Flat with spot & 5.22 & 1.31 & 15.4 & 1.71 \\
Flat without spot & 5.98 & 1.49 & 11.5 & 1.28 \\
$100 \times$Solar with spot & 4.93 & 1.23 & 25.6 & 2.84 \\
$100 \times$Solar without spot & 5.66 & 1.42 & 17.2 & 1.91\\\hline
\end{tabular}
\end{table}

\subsection{Analysis on the transit depth variability}\label{sec:ap-depvar}
We analysed the transit depth variability to see if the TLSE causes the variability at the different stellar rotational phases. 

Transit analysis was performed following the method described in Section~\ref{sec:an-transit}. However, in this analysis, the transit depth was set as a separate parameter for each observation night, while the other transit parameters were common. Fitting was done for the four MuSCAT bands simultaneously. To reduce the computational cost, the GP hyperparameters were fixed with the best-fit values from the joint-fit results (Table~\ref{tab:trfit-results-eccentric}). Therefore the total number of parameters was 36: $T_0, P, \sqrt{e}\cos \omega, \sqrt{e}\sin \omega, a/R_\star, b$, and $(R_p/R_\star)_m$ for each light curve $m=\{1, 2, \cdots 22\}$, and $(q_{1B}, q_{2B})$ for each band $B=\{g, r, i, z\}$. Note that, since TESS transit data are not very high precision, we did not use them to examine transit depth variability in the first place. 

\begin{figure}
  \centering
      \includegraphics[width=0.45\textwidth]{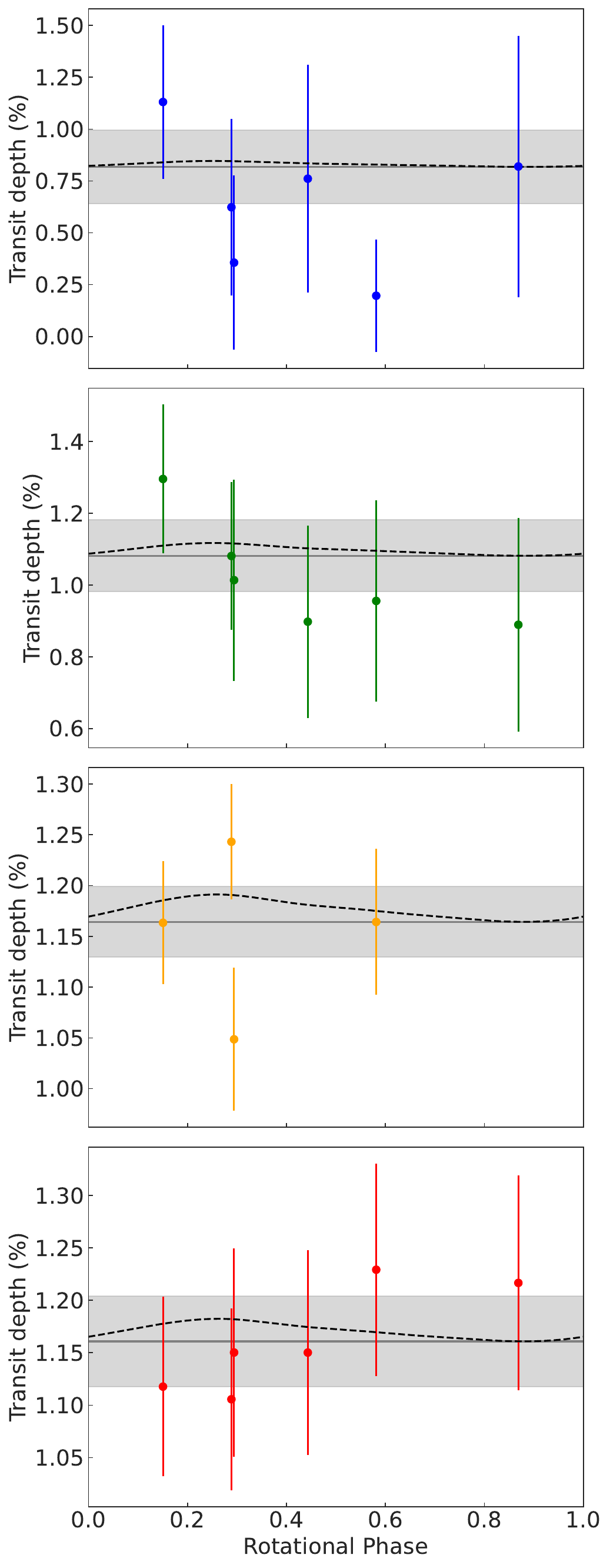}
  \caption{Derived transit depths according to the stellar rotation phase for $g$, $r$, $i$, and $z$-band from the top to the bottom. The grey lines and regions show the transit depths and their uncertainties derived by joint fit. The black dashed lines show the predicted transit depth variation from the TLSE. The derived uncertainties in the transit depths are larger than the predicted transit depth variation.}
  \label{fig:transit_depths_eachcolor}
 \end{figure}

Figure~\ref{fig:transit_depths_eachcolor} shows the derived transit depth variations in each band, along with the observing time and corresponding stellar rotational phase. The stellar rotational phase at each transit timing was calculated using the values of $P_{\rm rot}=1.87708$ day, described in Section~\ref{sec:an-rotation}. 

At the same time, we calculated the degree of TLSE (contamination factor) $\epsilon_B$ at each band $B$ per stellar rotation phase. In principle, the derivation of $\epsilon_\lambda$ (Equation~\ref{eq:tlse-contrast}) means the ratio of the `flux from the spotless stellar surface' to the `flux from the current stellar surface'. This means that we can calculate the $\epsilon_B$ at each phase by dividing each phase-folded light curve by the maximum of the light curve. The dashed lines in Figure~\ref{fig:transit_depths_eachcolor} indicate the transit depth variation expected from the TLSE. The predicted differences in transit depths are only $ < 500$ ppm. This difference is not detectable with MuSCAT2/3 considering their photometric precision. This result was consistent with the result that transit depth was constant within the uncertainties.

\subsection{Analysis on the spot crossing probability}\label{sec:an-crossing}
We investigate the possibility of spot-crossing events observed, as it gives useful information to constrain the spot characteristics. We did not visually detect any signal of large spot crossings from the shape of transit light curves, other than the suspicious signal observed in observation $N_T=314$, which could not be determined whether it was spot-crossing or flare during transit. In another case, if there are many small spots on the transit chord, there should be random occurrences of those spots crossing during transits, which result in an increase in scatters during the transits when the light curves are stacked.

\begin{figure*}
  \centering
      \includegraphics[width=1.0\textwidth]{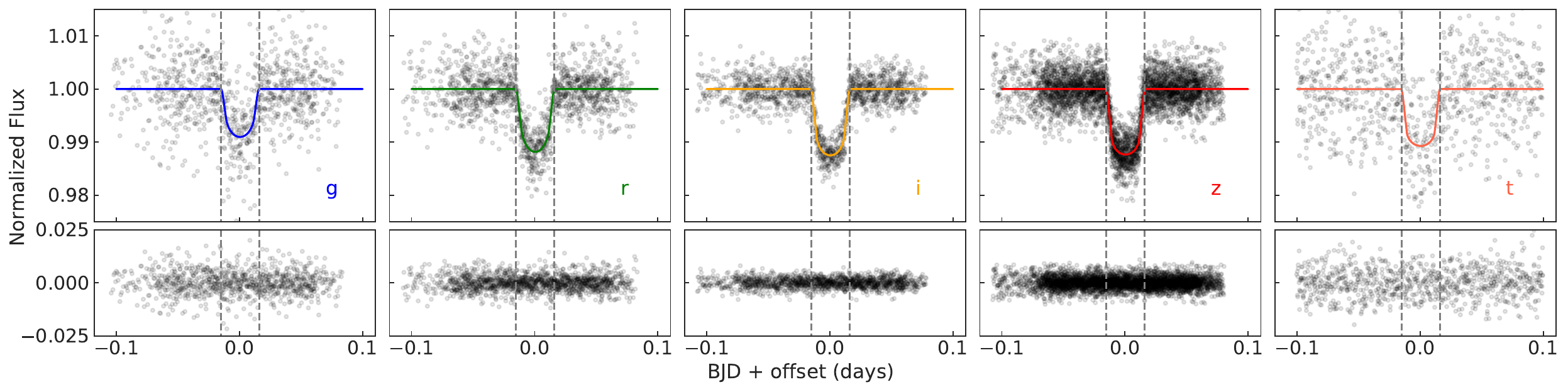}
  \caption{(Top) Best-fit transit models and observed data points phase-folded at the planetary orbital period after correcting baseline modulation. (Bottom) Residuals after subtracting the transit and baseline models from the observed data. There are no signs of spot crossings that would cause a larger RMS in transit.}
  \label{fig:whole_transit_stacked}
 \end{figure*}

Therefore, we stacked the light curves in each band after GP detrending, as shown in Figure~\ref{fig:whole_transit_stacked}. We also removed the transit signal from the light curves and calculated the RMS values of the residual, both in-transit and out-of-transit. The calculated RMS values are summarized in Table~\ref{tab:rms}. There is no significant increase in RMS in transit, indicating that there is no strong evidence of spot crossings.

The result is consistent with the fact that none of the transit observations by \citet{Thao2020} and \citet{Kain2020} gave a clear signal of spot crossings; one of the light curves by the Spitzer showed a signal that would be spot crossing, but the possibility of systematic error remains. These results suggest that the spots of K2-25 are not very large or not located at the transit chord.

\begin{table}
\centering
\caption{Calculated RMS for the phase-stacked light curves in each band, after transit+baseline model correction.}\label{tab:rms}
\begin{tabular}{c|cc}\\
       & \multicolumn{2}{c}{RMS}     \\
Band   & in-transit & out-of-transit \\ \hline
$g$    & 0.00590    &  0.00555      \\
$r$    &  0.00347   &   0.00370             \\
$i$    &   0.00218         &     0.00238           \\
$z$  &   0.00278         &      0.00303          \\
TESS &    0.00599        &       0.00691         \\ \hline
\end{tabular}
\end{table}

\subsection{Analysis on the flares from TESS light curve} \label{sec:an-flare}
From visual inspection, we found there are several flare-like signals in the TESS light curve. Frequent flares would not only lead to errors in transit depth measurements but would be another source of transit depth variations.

Therefore, we derived the flare frequency (how often flares of each energy occur) of the system, following the five steps; (1) detrending the light curves, (2) searching for flare candidates, (3) fitting flare models to the candidate light curve, (4) deriving total energy for each flare, and (5) deriving flare frequency.

First, we detrended the light curve. Assuming that the flares occur almost independently of the stellar rotation phase, models fitted to light curves folded in the stellar rotation phase can be regarded as unaffected by flares. Therefore, we detrended the TESS light curve by the phase-folded GP model described in Section~\ref{sec:an-amplitude} and unfolded it with time as the horizontal axis.

\begin{figure*}
    \centering
    \includegraphics[width=0.9\textwidth]{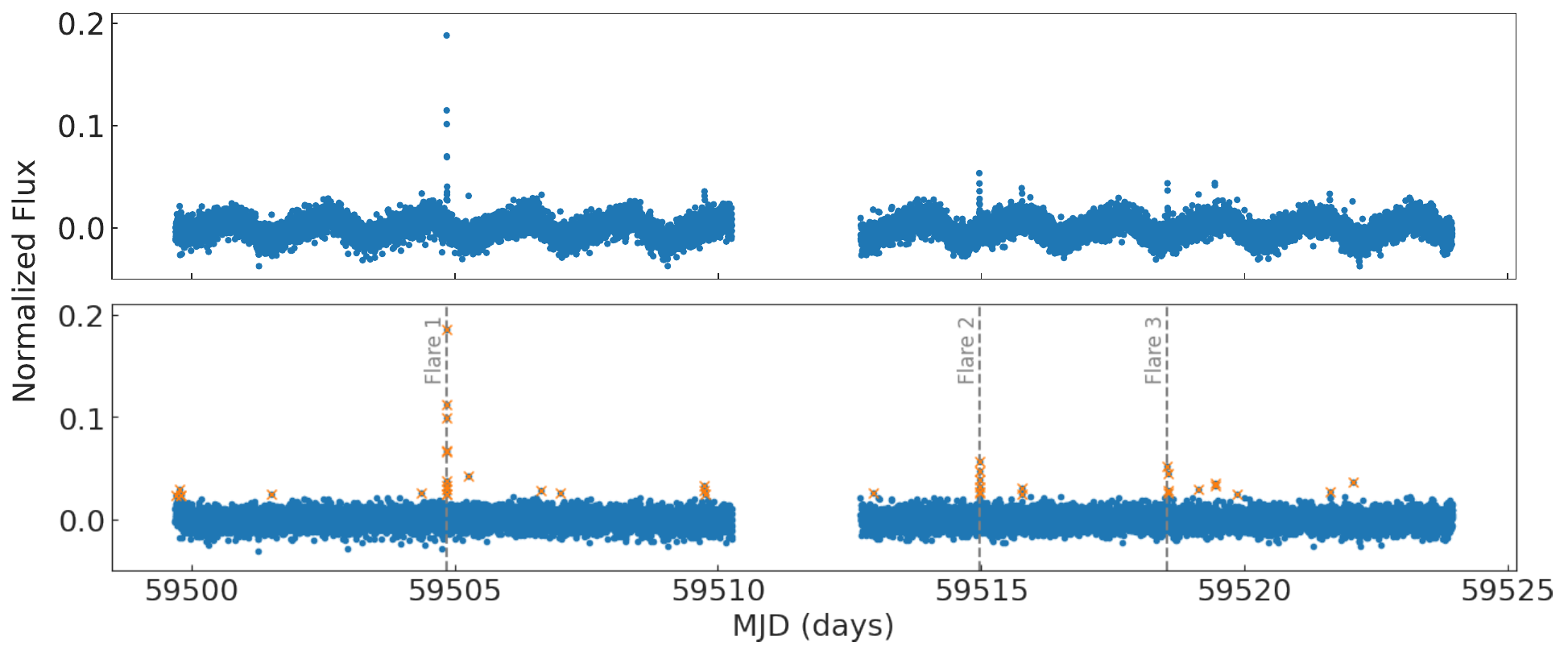}
    \caption{Normalized \texttt{PDCSAP} light curve of K2-25 from TESS Sector 44 (top) and the detrended light curve by phase-folded GP model (bottom). Data points whose normalized flux values are more than $3 \sigma$ away from the median value of the entire detrended curve are marked with orange crosses. The flare candidates are indicated by the grey dotted lines.} 
    \label{fig:flare_detection}
\end{figure*}

Second, the flare candidates were searched using the criteria described in \citet{Chang2015}: ``There are three or more consecutive points where the flux is more than $3\sigma$ off the overall median value". As a result, three flare candidates were detected. We also visually checked all the other data points that excess $3\sigma$ from the median and did not confirm flare-like signals other than these three. Figure~\ref{fig:flare_detection} shows the normalized TESS light curves, with and without detrending, along with the times of three detected flare candidates.

We cut out a light curve at 0.02 days before and 0.08 days after the first rise of each flare-like signal. To model the flare light curves, we used the empirical flare template \texttt{Llamaradas Estelares} \citep{Mendoza2022}, which is an updated model from \citet{Davenport2014}. The flare shape in the light curves $F_{\rm flare}(t)$ is characterized by three parameters: the time of flare peak $t_{\rm peak}$, the time-scale of decay FWHM, and the peak amplitude of the flux $f_{\rm amp}$. Independently for each flare candidate, parameter estimation was performed by \emcee to maximise the likelihood of the flare model to the observed flare light curve. The \emcee chain was well-converged, indicating that the probability that these are typical flares is high.

\begin{figure*}
    \centering
    \includegraphics[width=0.85\textwidth]{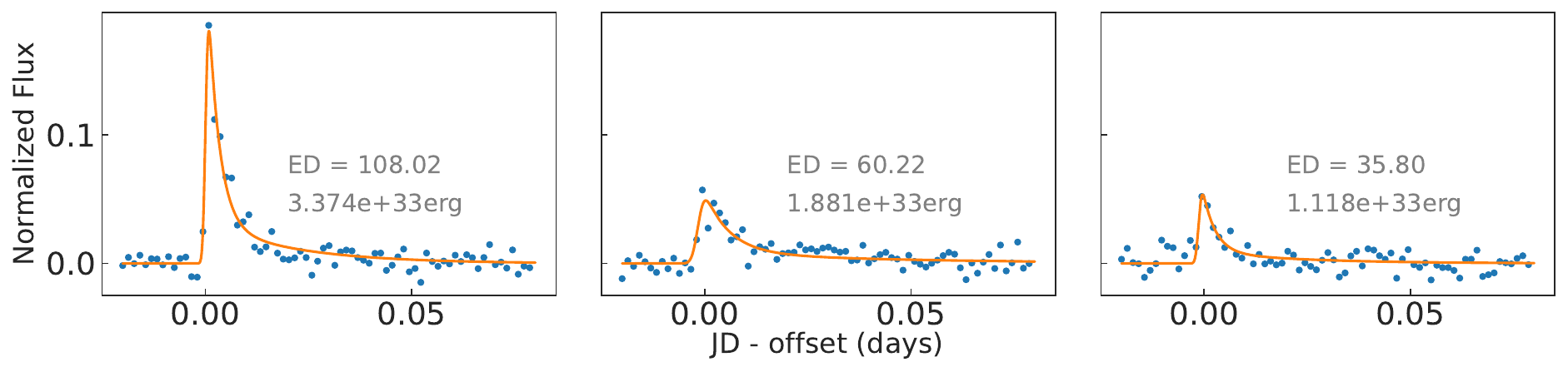}
    \caption{Close-up of the light curves of visually selected flare candidates (flare 1 to 3, from left to right). The orange line shows the best-fit flare models. Calculated equivalent duration (ED) and flare energy are shown next to the models.} \label{fig:flare_fit}
\end{figure*}

Therefore, the equivalent duration (ED) of the flare and flare energy at the observed band $E_{\rm flare}$ were calculated from the flare flux variation $F_{\rm flare}(t)$, by the following equations described in \citet{Hunt-Walker2012}.
\begin{align}
    {\rm ED} &= \int \frac{F_{\rm flare}(t)}{F_0}dt,\\
    E_{\rm flare} &= L_\star \times {\rm ED},
\end{align}
where $F_0$ is the quiescent stellar flux and $L_\star$ is the stellar luminosity in the observed band. We calculated ED and $E_{\rm flare}$ from $F_{\rm flare}(t) / F_0 $ as the best-fit normalized model and $L_\star = 0.00816 L_\odot = 3.12\times10^{31}$ erg/sec from \citet{Thao2020}. Figure~\ref{fig:flare_fit} shows the light curves of three detected flares with the model curve, and calculated ED and $E_{\rm flare}$. Calculated flare energy values are $\sim 3\times10^{33}$ erg, $2\times10^{33}$ erg, and $1\times10^{33}$ erg for three flares, respectively.
 
Although the number of flares detected is small, flare frequency was roughly estimated. The observation period of the TESS light curve was $21.76$ days, not including the gap for data downlink. Within this observing window, we found three flares with the energy at $\sim 10^{33}$\,erg and did not find flares with higher energy. We estimate the $\sim 10^{33}$ erg-class flare happens $\sim 0.1$ per day. The smaller flares should happen more frequently, but they seem to be below the detection limit ($\sim 10^{32}$~erg).

\begin{figure*}
  \centering
  \includegraphics[width=0.8\textwidth]{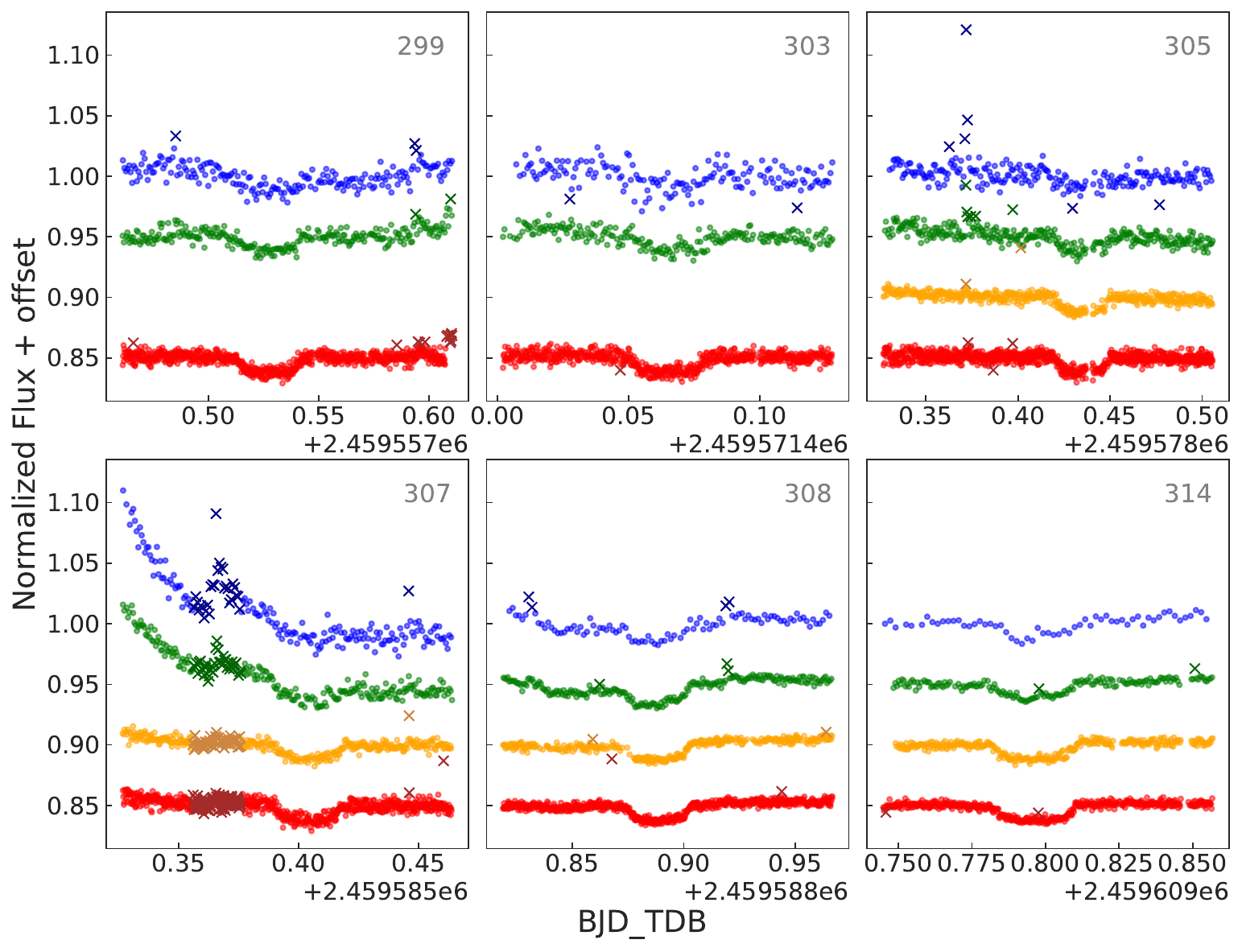}
  \caption{Raw light curves by MuSCAT2/3 before removing outliers, in $g$, $r$, $i$, $z$-band from top to bottom at each panel, at each night. The coloured dots indicate the data points used for transit analysis, and cross marks indicate the removed data points. The values at the upper-right corner of each panel indicate the $N_{\rm T}$ values. There are several flare-like signals detected and removed as outliers. The details of outlier removal are described in Section~\ref{sec:muscats}.}
  \label{fig:outliers}
 \end{figure*}

Such flares can affect the accurate measurement of transit depths. Indeed, our transit light curves by MuSCAT2/3 showed flare-like brightening during the transit observations (Figure~\ref{fig:outliers}). We did not model the flare in the transit analysis, as the observational precision is not that high compared to the flare amplitude. The influence of the flare on the transit depth measurement should have been suppressed by removing outliers and using a Gaussian process as a noise model. However, in the future, when more precise transit observations are made by space telescopes, attention should be paid to whether flares are occurring during, before, or after the transit.

\subsection{Analysis on the effectiveness of multi-band monitoring} \label{sec:ap-multi}

To confirm the effectiveness of multi-band monitoring observations for spot characteristics derivation, we tried the spot mapping analysis described in Section~\ref{sec:an-distribution} only using the TESS light curve, to compare with the result using all five light curves. 

\begin{figure*}
  \centering
      \includegraphics[width=0.95\textwidth]{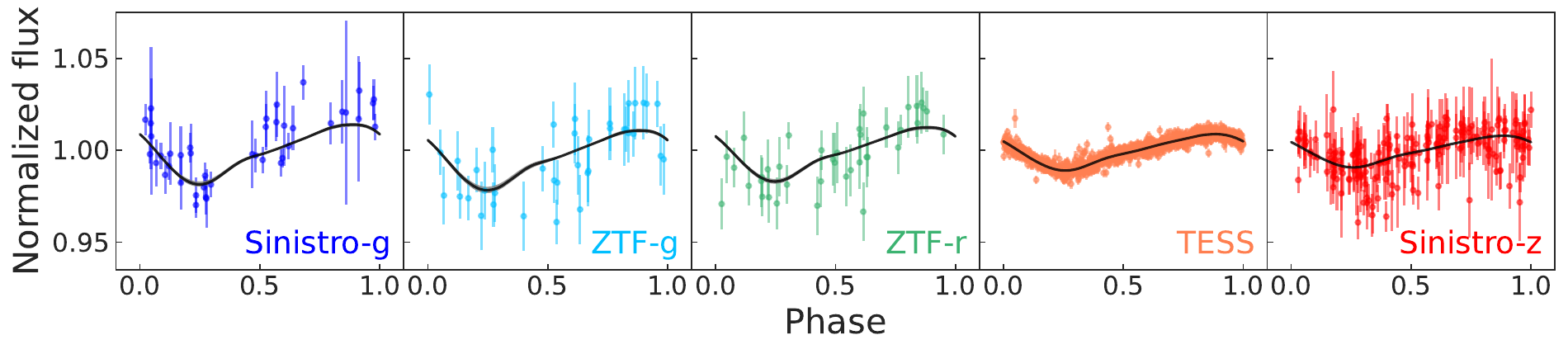}
      \includegraphics[width=0.95\textwidth]{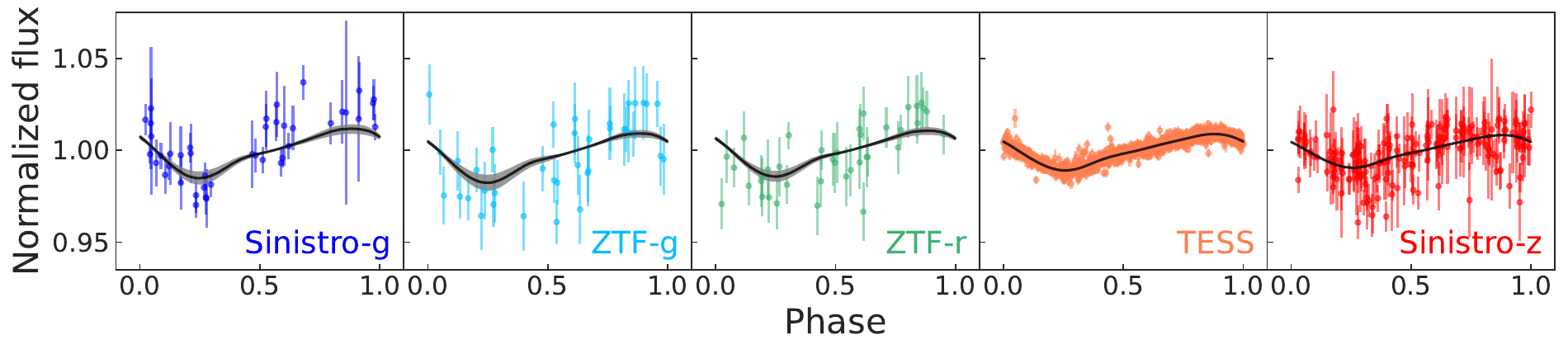}
  \caption{\textit{(Top.)} The phase-folded light curves and fitted model curves derived by the spot mapping with the three-spot case. The solid line shows the best-fit curves and the shaded region shows a $1 \sigma$ deviation of model uncertainties, derived from the models drawn from the converged MCMC chain of the spot parameters.  \textit{(Bottom.)} As a comparison, a result of model fitting when only TESS light curves are used. Light curves are more well-constrained when we use multi-band light curves.}
  \label{fig:effective_multi}
 \end{figure*}

\begin{figure*}
  \centering
      \includegraphics[width=0.95\textwidth]{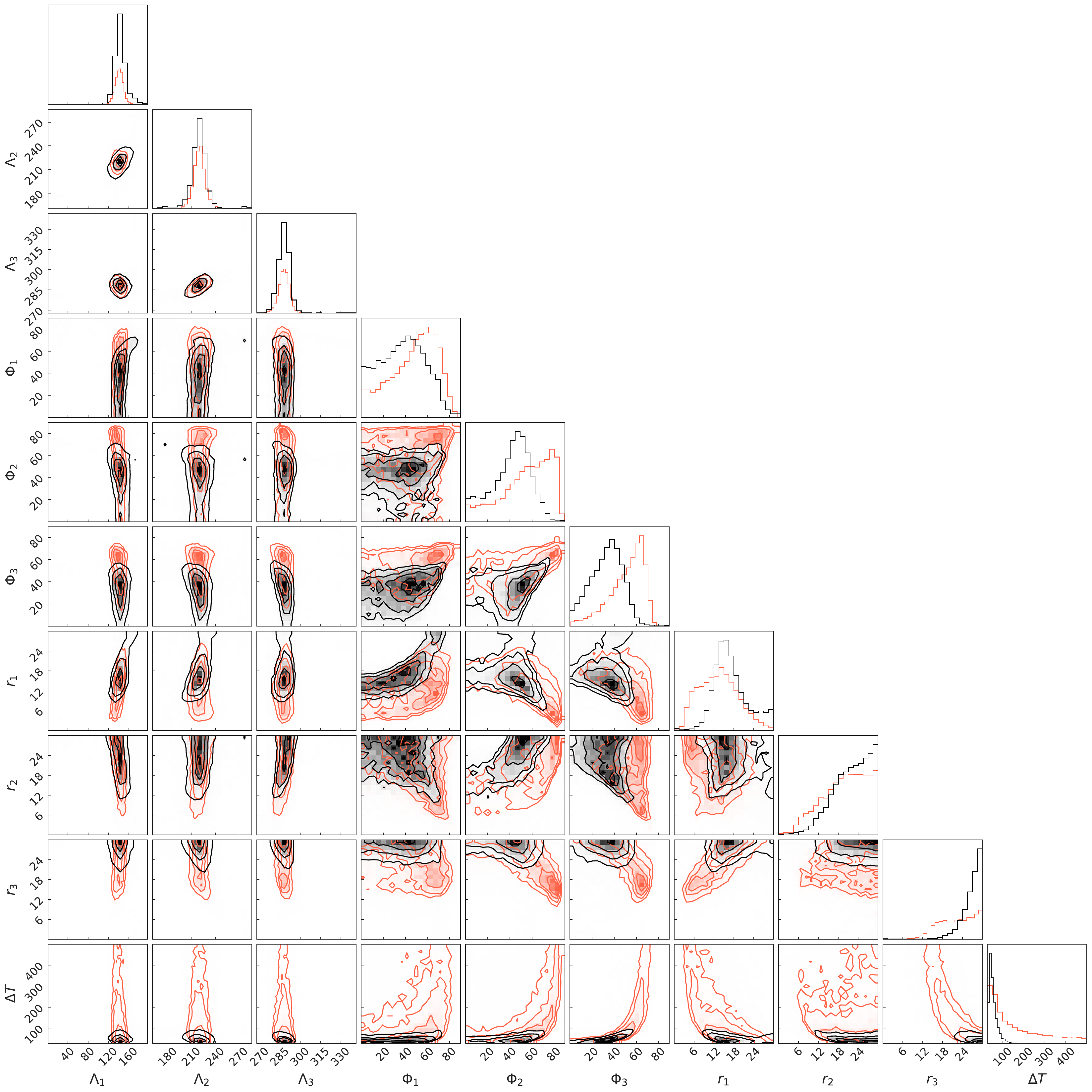}
  \caption{Corner plot to show the posterior distribution of parameters for spot mapping using all five light curves (black) overlayed with the posteriors using only TESS light curve (orange); spot longitude $\Lambda_n$, latitude $\Phi_n$ and spot radius $r_n$ for spot $n=\{1, 2, 3\}$ with three spot assumption. In the multi-band case, the spot latitudes are more limited not to reach higher values, thereby the spot temperature is more strongly constrained. } 
  \label{fig:corners_all}
 \end{figure*}

Figure~\ref{fig:effective_multi} compares the resulted best-fit light curve models and their $1\sigma$ uncertainties, and Figure~\ref{fig:corners_all} shows the posterior distribution of spot parameters for a three-spot model, which compares the result of TESS-only light curve v.s. five-band light curves. When only the TESS light curve was used, the MCMC did not converge with the reasonable iteration steps. Therefore, we needed to put additional Gaussian prior for three parameters that describe the spot longitudes, and a uniform prior for $\Delta T$ to take values between 0 to 500\,K, for the TESS-only case. Nevertheless, the five-band light curve case resulted in more strict constraints on the models and posterior distribution of spot parameters. Especially, while the TESS-only case allows spot latitude to be high (intuitively, to create a more smoothly modulating light curve),  the spot latitudes are more constrained to be $\sim 40$ deg in the five-band light curve case. This resulted in a better constraint on spot temperature, by reflecting the correlations between spot parameters. The $2\sigma$ upper limit of $\Delta T$ was 407\,K when only the TESS light curve is used, compared with the 86\,K when five-band light curves are used. The results show that the light curves from ground-based telescopes, which have lower time coverage and photometric precision compared to the TESS light curve, still work toward reducing the model uncertainties.

\bsp	
\label{lastpage}
\end{document}